\documentstyle[aps,floats,prd,psfig]{revtex}

\newcommand{\ApJL}{Astrophys. J. Lett.}
\newcommand{\ApJ}{Astrophys. J.}
\newcommand{\PRL}{Phys. Rev. Lett.}
\newcommand{\PRD}{Phys. Rev. D}
\newcommand{\MNRAS}{MNRAS}
\newcommand{\AsAs}{A\&A}
\newcommand{\aut}[2]{{#2.\ #1}}
\newcommand{\refs}[6]{#2, {\bf #3} {#4} (#5)}

\newcommand{\amp}{and }
\newcommand{\spin}[1]{\,{}_{#1}^{\vphantom{m}}}   
\renewcommand{\l}{{\bf l}}
\newcommand{\la}{{{\bf l}_1}}

\newcommand{\da}{d_A}

\newcommand{\tot}{{\rm t}}

\newcommand{\cmb}{\Theta}
\newcommand{\s}{{\rm s}}
\newcommand{\n}{{\rm n}}
\newcommand{\vecl}{{\bf l}}
\newcommand{\vecla}{{{\bf l}_1}}
\newcommand{\veclb}{{{\bf l}_2}}
\newcommand{\veclc}{{{\bf l}_3}}
\newcommand{\vecld}{{{\bf l}_4}}
\newcommand{\intl}[1]{\int {d^2 {\bf l}_#1 \over (2\pi)^2}}

\newcommand{\bfl}{{\mathbf{l}}}
\newcommand{\dirac}{{\rm D}}

\newcommand{\lens}{{\rm len}}
\newcommand{\pp}{{\phi\phi}}

\newlength{\tskip}
\newlength{\colwidth}

\newcommand{\beq}{\begin{equation}}
\newcommand{\eeq}{\end{equation}}
\newcommand{\beqa}{\begin{eqnarray}}
\newcommand{\eeqa}{\end{eqnarray}}

\newcommand{\deld}{\delta^{\rm D}}
\newcommand{\bn}{\hat{\bf n}}
\newcommand{\bm}{\hat{\bf m}}

\newcommand{\rad}{r}    

\newcommand{\len}{\phi}

\newcommand{\sz}{{\rm SZ}}

\begin{document}

\title{Weak Lensing of the CMB: Extraction of Lensing Information from the Trispectrum}
\author{Asantha Cooray and Michael Kesden}
\address{
Theoretical Astrophysics, California Institute of Technology,
Pasadena, California 91125\\}


\maketitle


\begin{abstract}
We discuss the four-point correlation function, or the trispectrum in Fourier
space, 
of CMB temperature and polarization anisotropies due to the weak gravitational
lensing effect by intervening large scale structure.
We discuss the squared temperature power spectrum as a probe of this trispectrum and, more importantly, as an
observational approach to extracting the power spectrum of the deflection angle
associated with the weak gravitational lensing effect on the CMB. We extend
previous discussions on the trispectrum and associated weak lensing
reconstruction from CMB data by calculating 
non-Gaussian noise contributions, beyond the previously discussed dominant
Gaussian noise.  Non-Gaussian noise contributions are generated by lensing
itself and by the correlation between the
lensing effect and other foreground secondary anisotropies in the CMB such as
the Sunyaev-Zel'dovich (SZ) effect. 
When the SZ effect is removed from temperature maps using its
spectral dependence, we find these additional non-Gaussian noise contributions
to be an order of magnitude lower than
the dominant Gaussian noise. If the noise-bias due to the dominant Gaussian part of the
temperature squared power spectrum is removed, then these additional
non-Gaussian contributions provide the limiting noise level for the lensing reconstruction.
The temperature squared power spectrum allows a high signal-to-noise
extraction of the lensing deflections and a 
confusion-free separation of the curl (or B-mode) polarization due to
inflationary gravitational waves from that due to lensed gradient (or E-mode)
polarization.
The small angular scale temperature and polarization anisotropy measurements provide a novel  approach to
weak lensing studies, complementing the approach based on galaxy ellipticities.
\end{abstract}





\section{Introduction}

The well understood features in the angular power spectrum of cosmic microwave background (CMB) anisotropies,
such as acoustic peaks and the damping tail \cite{PeeYu70}, allow a useful probe of the cosmology
\cite{Jugetal95}. The ability of CMB anisotropies to constrain most, or certain combinations of, 
parameters  that define cold dark matter cosmological models with adiabatic initial conditions 
has driven a significant number of experiments from ground and space. These experiments include
NASA's MAP mission\footnote{http://map.nasa.gsfc.gov} and, in the long term,
ESA's Planck surveyor\footnote{http://astro.estec.esa.nl/Planck/; also, ESA D/SCI(6)3.}.
With the successful detection of the first two or three acoustic peaks at
degree angular scales \cite{Miletal99}, 
the anisotropy experiments have started to focus on fluctuations on
angular scales of arcminutes or less. 
At these angular scales, fluctuations are mostly dominated
by secondary effects due to the local large scale structure (LSS) between us
and the last scattering surface \cite{Coo02}. 
In addition to generating new anisotropies via scattering and gravitational
infall,
non-linear effects such as the weak gravitational lensing of the
CMB \cite{Blaetal87} leave important imprints that can in turn be used to
probe cosmology or 
astrophysics related to the evolution and growth of structure (e.g., \cite{SpeGol99,Zal00,Hu01,Benetal00}).

The lensing effect on the CMB leads to modifications in the direction of photon
propagation while leaving the 
surface brightness of the CMB unaffected, implying that its contributions are
effectively at the second order level in
temperature fluctuations. This second order contribution is probed by the
three-point 
correlation function of the CMB \cite{SpeGol99}.  Here, contributions arise
mainly due to the fact that potentials
 that deflect CMB photons also correlate with other secondary effects from the
foreground
large scale structure, such as the integrated Sachs-Wolfe (ISW; \cite{SacWol67}) effect or the Sunyaev-Zel'dovich
(SZ; \cite{SunZel80}) effect. At the four-point level, the weak lensing effect
itself contributes to the correlation function
due to its non-linear mode-coupling nature. The trispectrum, the four-point
correlation function in Fourier space,
 due to lensing was discussed in \cite{Zal00} and the same trispectrum, under
an all-sky formulation, is presented in \cite{Hu01}.
In \cite{Coo02b}, we presented the trispectrum due to lensing alone and the
correlation between
lensing and secondary effects and focused on the resulting non-Gaussian
covariance of the power spectrum measurements of the CMB temperature
fluctuations.

Here, we study an important application of the CMB trispectrum involving a
model independent recovery of statistics related to the
foreground angular deflections of CMB photons
(or alternatively the projected potentials that generate these deflections).
In terms of the trispectrum formulated under a flat-sky approximation, we show explicitly
how one can use a higher order statistic such as the power spectrum of squared
temperatures to extract the lensing information. The approach described in the
present paper 
was first introduced by Ref. \cite{Hu01} while other approaches were considered
in Refs. \cite{SelZal99,Zal00,Benetal00}
based on CMB temperature and polarization observations.
We discuss the full trispectrum formed by the weak lensing effect on the CMB
and
extend previous work on this subject by calculating in detail the
noise associated with lensing extraction. 
We also discuss noise contributions resulting from the CMB trispectrum formed
via the correlation between lensing and 
secondary effects due to the large scale structure. Since the lensing effect
modifies the damping tail of CMB anisotropies 
and the information from the damping tail is used to reconstruct lensing
statistics, we suggest that any secondary
effect that dominates small angular scale fluctuations can have a significant
impact on the lensing extraction.
An important contribution at these scales is the SZ effect; fortunately this
contribution can be removed from thermal CMB maps by noting its distinct
spectral signature in observations involving many frequencies about the SZ
null at $\sim$ 217 GHz \cite{Cooetal00}.

As discussed in \cite{Kesetal02}, the lensing extraction is essential for an
important application utilizing CMB polarization observations. 
It is now well known that the curl (or B-modes) of the polarization 
allows a direct detection of the gravitational wave contribution
\cite{KamKosSte97}. 
The amplitude of the angular power spectrum of this polarization signal is
proportional to the square of the
energy scale of inflation, and thus any detection of the B-mode polarization
allows a probe of the inflationary energy scale.
There is one significant source of confusion, however, in the B-mode map which
can potentially limit the detection of the
gravitational wave signal. As discussed in Ref. \cite{ZalSel98}, the weak
lensing effect on CMB polarization results in
a fractional conversion of the dominant E-mode polarization due to density
perturbations to the B-mode.
The contribution from lensing to the B-mode map is such that its power spectrum
is at the level of the polarization signal expected
from gravitational waves if the energy scale of inflation is at scales of
$\sim$ $10^{16}$ GeV.  Since we do not know the energy scale
of inflation a priori, if it is sufficiently small the lensing effect will
easily dominate the contribution from primordial gravitational waves.

In order to identify the gravitational wave signature with no confusion, one
must separate the contribution due to lensing
reliably. As discussed in \cite{Kesetal02}, observations of weak lensing in
large scale structure via galaxy ellipticities
do not allow a useful separation of the lensing contribution.  A substantial
contribution to the effect comes from
potentials at redshift greater than 1 to 2, the redshift range currently
accessible with galaxy lensing surveys. Even if
we have a reliable lensing survey out to a redshift of 5, we will only be able
to reduce the lensing signal by
a factor of $\sim$ 4, which is hardly of any use given the order of magnitude
change in the gravitational wave contribution that is possible if the
inflationary energy scale
is much less than that related to grand unified theories \cite{KamKos99}. Thus,
a better approach is to
use a source at the last scattering itself, ie. CMB anisotropies,
to reconstruct the lensing potential. With such a reconstruction, one can
clean the polarization maps and
obtain unbiased and model independent estimates of the polarization signal at
the last scattering surface.
Here, we discuss in detail the approach related to the CMB trispectrum and
associated non-Gaussian noises, ignored previously in
lensing reconstruction \cite{Hu01b,HuOka01} and in the application of lensing
reconstruction to separate the gravitational wave
signature \cite{Kesetal02}. Under certain conditions, we find that these
additional noise contributions are an 
order of magnitude below the dominant Gaussian noise contribution. If the
noise-bias associated with the
Gaussian part of the temperature squared power spectrum is removed, then the
limiting noise level for lensing extraction is determined
by the non-Gaussian contribution presented here.  Even if the noise-bias is
not removed, we note that the proposed statistic
allows a high signal-to-noise extraction of the lensing signal and a
separation of the confusing lensed E-mode contribution in the
B-mode map from those due to inflationary gravitational waves. 

In \S~\ref{sec:generallensing}, we introduce the basic
ingredients for the present calculation. The CMB anisotropy trispectra due to
weak lensing and correlations
between weak lensing and secondary effects are derived in
\S~\ref{sec:lensing}.  In \S~\ref{sec:covariance}, we introduce a statistic
involving the power
spectrum of squared temperatures, instead of the usual temperature itself, as
a probe of the angular
power spectrum of lensing deflections and extend the discussion to involve
polarization in
\S~\ref{sec:EBtri}. In \S~\ref{sec:results}, we
discuss our results and conclude with a summary in \S~\ref{sec:summary}. The
Appendix contains useful formulae
related to additional estimators of lensing based on polarization and a
combination of temperature and polarization.

\section{Gravitational Lensing}
\label{sec:generallensing}

Large-scale structure between us and the last scattering surface deflects 
CMB photons propagating towards us. Since the lensing effect on CMB 
is essentially a redistribution of photons from large scales to small 
scales, the resulting effect appears only in the second order \cite{Hu00}. 
In weak gravitational lensing, the deflection angle on the sky is given by 
the angular gradient of the lensing potential, 
$\alpha(\bn) = \nabla \phi(\bn)$, which itself is a 
projection of the gravitational potential, $\Phi$ (see e.g. \cite{Kai92}),
\begin{eqnarray}
\phi(\bm)&=&- 2 \int_0^{\rad_0} d\rad 
\frac{\da(\rad_0-\rad)}{\da(\rad)\da(\rad_0)}                
\Phi (\rad,\hat{{\bf m}}\rad ) \,.
\label{eqn:lenspotential}
\end{eqnarray}

The quantities here are the conformal distance or 
lookback time, from the observer, given by 
\begin{equation}
\rad(z) = \int_0^z {dz' \over H(z')} \,,
\end{equation}
and the analogous angular diameter distance
\begin{equation}
\da = H_0^{-1} \Omega_K^{-1/2} \sinh (H_0 \Omega_K^{1/2} \rad)\,, 
\end{equation}
with the expansion rate for adiabatic CDM 
cosmological models with a cosmological constant given by
\begin{equation}
H^2 = H_0^2 \left[ \Omega_m(1+z)^3 + \Omega_K (1+z)^2              
+\Omega_\Lambda \right]\,.
\end{equation}
Here, $H_0$ can be written as the inverse
Hubble distance today $cH_0^{-1} = 2997.9h^{-1} $Mpc.
We follow the conventions that in units of the critical 
density $3H_0^2/8\pi G$, 
the contribution of each component is denoted 
$\Omega_i$, $i=c$ for the CDM, $b$ for the baryons, 
$\Lambda$ for the cosmological constant. 
We also define the auxiliary quantities 
$\Omega_m=\Omega_c+\Omega_b$ and $\Omega_K=1-\sum_i \Omega_i$, 
which represent the matter density and the contribution of 
spatial curvature to the expansion rate respectively. 
Note that as $\Omega_K \rightarrow 0$, 
$\da \rightarrow \rad$ and we define $\rad(z=\infty)=\rad_0$. 
Though we present a general derivation of the trispectrum contribution 
to the covariance, we show results for the currently 
favorable $\Lambda$CDM cosmology with 
$\Omega_b=0.05$, $\Omega_m=0.35$, $\Omega_\Lambda=0.65$ and $h=0.65$.

The lensing potential in equation~(\ref{eqn:lenspotential}) 
can be related to the well known convergence generally
 encountered in conventional lensing 
studies involving galaxy shear \cite{Kai92}
\begin{eqnarray}
\kappa(\bm) & =& {1 \over 2} \nabla^2 \phi(\bm) \nonumber \\           
& = &-\int_0^{\rad_0} d\rad \frac{\da(\rad)\da(\rad_0-\rad)}{\da(\rad_0)}\nabla_{\perp}^2 \Phi (\rad ,\hat{{\bf m}}\rad) \, , \nonumber \\
\end{eqnarray}
where note that the 2D Laplacian operating on 
$\Phi$ is a spatial and not an angular Laplacian.
Expanding the lensing potential to Fourier moments,
\begin{equation}
\phi(\bn) = \int \frac{d^2\vecl}{(2\pi)^2} \phi(\vecl) 
{\rm e}^{i \vecl \cdot \bn}  \, ,
\end{equation}
we can write the usually familiar quantities of convergence and shear components of 
weak lensing as \cite{Hu00}
\begin{eqnarray}
\kappa(\bn) &=& -\frac{1}{2}\int \frac{d^2\vecl}{(2\pi)^2} 
l^2 \phi(\vecl) {\rm e}^{i\vecl \cdot \bn} \nonumber \\
\gamma_1(\bn) \pm i\gamma_2(\bn) &=& -\frac{1}{2}\int \frac{d^2\vecl}{(2\pi)^2} 
l^2 \phi(\vecl) {\rm e}^{\pm i 2 (\varphi_l-\varphi)}{\rm e}^{i\vecl \cdot \bn} \, .
\label{eqn:kappa}
\end{eqnarray}
While the two terms $\kappa$ and $\phi$ contain  
differences with respect to radial and wavenumber weights,
we note that these differences are artificial and that they cancel under the Limber approximation \cite{Lim54}. 

The power spectrum of the lensing potential is defined as
\begin{equation}
\langle \phi(\vecl) \phi(\vecl') \rangle = (2\pi)^2 \delta_D(\vecl+\vecl') C_l^\pp \, ,
\end{equation}
where $\delta_D$ is the Dirac delta function.
Expanding the gravitational potential to density perturbations 
using the Poisson equation \cite{Bar80}
\begin{equation}
\Phi = {3 \over 2} \Omega_m \left({H_0 \over k}\right)^2        
\left( 1 +3{H_0^2\over k^2}\Omega_K \right)^{-1} \frac{G(r)}{a}      
  \delta(k,0)\,,
\label{eqn:Poisson}
\end{equation}
and the expansion of a plane wave,  
we can write the power spectrum of lensing potentials as
\begin{eqnarray}
C_l^{\phi\phi} &=& \frac{2}{\pi} \int k^2\, dk P(k)                
\int d\rad_1 W^\lens(k,\rad_1) j_l(k\rad_1) \int d\rad_2 W^\lens(k,\rad_2) j_l(k\rad_2) \, .
\label{eqn:cllens}
\end{eqnarray}
Here,
\begin{eqnarray}
W^\lens(k,r) &=& -3\Omega_m \left(\frac{H_0}{k}\right)^2 \frac{G(\rad)}{a} 
\frac{\da(\rad_0-\rad)}{\da(\rad)\da(\rad_0)} \, 
\end{eqnarray}
and we have introduced the power spectrum of density fluctuations
\begin{equation}
\left< \delta({\bf k})\delta({\bf k')} \right> = (2\pi)^3        
\deld({\bf k}+{\bf k'}) P(k)\,,
\end{equation}
where
\begin{equation}
\frac{k^3P(k)}{2\pi^2} = \delta_H^2 
\left({k \over H_0} \right)^{n+3}T^2(k) \,,
\end{equation}
in linear perturbation theory.  
We use the fitting formulae of \cite{EisHu99} 
in evaluating the transfer function $T(k)$ for CDM models. 
Here, $\delta_H$ is the amplitude of present-day density fluctuations 
at the Hubble scale; with $n=1$, we adopt the COBE normalization 
for $\delta_H$  of $4.2 \times 10^{-5}$ \cite{BunWhi97}, 
consistent with galaxy cluster abundance ($\sigma_8=0.86$ \cite{ViaLid99}), 
  Note that in linear theory, the power
spectrum can be scaled in time, $P(k,r)=G^2(r)P(k,0)$, using the the growth function  \cite{Pee80}
\begin{equation}
G(r) \propto {H(r) \over H_0} \int_{z(r)}^\infty dz' (1+z') 
\left( {H_0\over H(z')} \right)^3\,.
\end{equation}
In the non-linear regime, one can use prescriptions such as the fitting function by \cite{PeaDod96} 
to calculate the fully non-linear density field power spectrum.

Note that an expression of the type in 
equation~(\ref{eqn:cllens}) can be evaluated 
efficiently with the Limber approximation \cite{Lim54}.
Here, we employ a version based on
the completeness relation of spherical Bessel functions \cite{KamSpe94}
\begin{equation}
\int dk \, k^2 F(k) j_l(kr) j_l(kr')  \approx {\pi \over 2} \da^{-2} \deld(r-r')
                                                F(k)\big|_{k={l\over d_A}}\,,
\label{eqn:ovlimber}
\end{equation}
where the assumption is that $F(k)$ is a slowly-varying function. A similar approach is
also presented in \cite{Val00}.  Using this, we obtain a useful approximation for the 
power spectrum of lensing potentials as
\begin{equation}
C_l^\pp = \int_0^{\rad_0} \frac{d\rad}{\da^2}\, 
\left[W^\lens({l \over \da},r)\right]^2         
P\left({l \over\da};\rad\right) \, .
\label{eqn:powerform}
\end{equation}
Using equation~(\ref{eqn:kappa}), we note that
the power spectrum of convergence is related to that of 
the potentials through $C_l^\kappa = 1/4 l^4 C_l^\pp$.

\section{Lensing Contribution to CMB}
\label{sec:lensing}

In order to derive weak lensing contributions to the CMB trispectrum, we
follow  Hu in Ref. \cite{Hu00} and Zaldarriaga in Ref. \cite{Zal00}. We formulate the 
contribution under a flat sky approximation. In general, the flat-sky 
approach simplifies the derivation and  computation through the replacement of summations over Wigner symbols 
through integrals involving mode coupling angles.

As discussed in prior papers \cite{SpeGol99,Hu00},  weak lensing remaps temperature projected on the sky
through the angular deflections 
resulting along the
photon path by
\begin{eqnarray}
\tilde \cmb(\bn) & = &  \cmb(\bn + \nabla\len) \nonumber\\
        & = &
\cmb(\bn) + \nabla_i \len(\bn) \nabla^i \cmb(\bn) + {1 \over 2} \nabla_i \len(\bn) \nabla_j \len(\bn)
\nabla^{i}\nabla^{j} \cmb(\bn)
+ \ldots 
\end{eqnarray}
As expected for lensing, note that the remapping conserves the 
surface brightness distribution of CMB.
Here, $\cmb(\bn)$ is the unlensed primary  component of CMB, the contribution at the last scattering surface, 
and  $\tilde \cmb(\bn)$ is the contribution  affected by the gravitational lensing deflections during the transit.
It should be understood that in the presence of low redshift
 contributions to CMB resulting through large scale 
structure, the temperature fluctuations also include a secondary 
contribution which we denote by $\cmb^\s(\bn)$ and, in all cases, a noise component denoted by $\cmb^\n(\bn)$
due to instrumental and detector limitations. We denote the total contribution
involving these through $\cmb^\tot(\bn)$, such that, $\cmb^\tot(\bn) = \tilde \cmb(\bn)+ \cmb^s(\bn)+\cmb^\n(\bn)$.
Since weak lensing deflection angles also trace the  large scale structure at low redshifts, 
secondary effects which are first order in density or potential 
fluctuations also  correlate with the lensing deflection angle $\phi$. These secondary effects include the ISW and SZ
and are discussed in \cite{SpeGol99,Coo02b}.

Taking  the Fourier transform, as appropriate for a flat-sky, we write
\begin{eqnarray}
\tilde \cmb(\vecla)
&=& \int d \bn\, \tilde \cmb(\bn) e^{-i \vecla \cdot \bn} \nonumber\\
&=& \cmb(\vecla) - \intl{1'} \cmb(\vecla') L(\vecla,\vecla')\,,
\label{eqn:thetal}
\end{eqnarray}
where
\begin{eqnarray}
\label{eqn:lfactor}
L(\vecla,\vecla') &=& \len(\vecla-\vecla') \, (\vecla - \vecla') \cdot \vecla'
+{1 \over 2} \intl{1''} \len(\vecla'') \\ &&\quad
\times \len^*(\vecla'' + \vecla' - \vecla) \, (\vecla'' \cdot \vecla')
                (\vecla'' + \vecla' - \vecla)\cdot
                             \vecla' \,.  \nonumber
\end{eqnarray}

We define the power spectrum and the trispectrum of the $\cmb^i$ fluctuation field, in the flat
sky approximation, following the usual way
\begin{eqnarray}
\left< \cmb^i(\bfl_1) \cmb^i(\bfl_2)\right> &=&
        (2\pi)^2 \delta_\dirac(\vecl_1+\vecl_2)  C_l^i\,,\nonumber\\
\left< \cmb^i(\bfl_1) \ldots
       \cmb^i(\bfl_4)\right>_c &=& (2\pi)^2 \delta_\dirac(\vecl_1+\vecl_2+\vecl_3+\vecl_4)
        T^i(\bfl_1,\bfl_2,\bfl_3,\bfl_4)\,, \nonumber \\
\end{eqnarray}
where the subscript $c$ denotes the connected part of the above four-point function and the
superscript $i$ denotes all possibilities including secondary anisotropies and noise.
We make the assumption that primordial fluctuations at the last scattering surface is Gaussian,
such that statistics are fully described by a power spectrum and the non-Gaussian four-point correlator,
$\left< \cmb(\bfl_1) \ldots \cmb(\bfl_4)\right>_c$ is zero. A non-Gaussian contribution is generated by
lensing effect and other secondary contributions.

\subsection{Power spectrum and Trispectrum}

The lensed power spectrum, according to the present formulation, is discussed in  \cite{Hu00} and we can write
\begin{eqnarray}
\tilde C_l^\cmb &=& \left[ 1 - \intl{1}
C^{\phi\phi}_{l_1} \left(\vecl_1 \cdot \vecl\right)^2 \right]	\, 
 	                        C_l^\cmb
        + \intl{1} C_{| \vecl - \vecl_1|}^\cmb C^{\phi\phi}_{l_1}
                [(\vecl - \vecl_1)\cdot \vecl_1]^2  \, .
\label{eqn:ttflat}
\end{eqnarray}

Thus, the total power spectrum of CMB anisotropies is
\begin{equation}
C_l^\tot = \tilde C_l^\cmb + C_l^\n +C_l^\s \, ,
\end{equation}
where we denote the power spectrum of the noise component by
\begin{eqnarray} \label{E:noise}
C_l^{\n} = f_{\rm sky} w^{-1} e^{l^2 \sigma_{b}^2} \, ,
\end{eqnarray}
when $f_{\rm sky}$ is the fraction of sky surveyed, $w^{-1} = 4\pi (s/T_{\rm
CMB})^2/(t_{\rm pix}N_{\rm pix})$ \cite{Kno95} is the variance per unit area
on the sky when $t_{\rm pix}$ is the time spent on
each of $N_{\rm pix}$ pixels with detectors of NET $s$, and $\sigma_b$ is the
effective beam-width of the instrument. The $C_l^\s$ represent the power spectrum associated with contributing
secondary effects. In the absence of secondary effects, especially in the cases where secondary contributions can
be separated from the primary fluctuations using statistical signatures or distinct information, such as the
spectral dependence of the SZ, we set $C_l^{\s}=0$. Note that in general, however, important secondary contributions such
as the ISW cannot be separated and will lead to additional noise contributions due to correlations with the lensing
potentials.

The calculation of the trispectrum follows similar to the power spectrum. 
Here, we explicitly show the calculation for  one term of the trispectrum and add all other terms through permutations.
First we consider the cumulants involving four temperature terms in Fourier
space:
\begin{eqnarray}
\left< \tilde \cmb(\bfl_1) \ldots
       \tilde \cmb(\bfl_4)\right>_c &=& \Big< \left( \cmb(\vecla) - \intl{1'} \cmb(\vecla')
L(\vecla,\vecla')\right) \left(\cmb(\veclb) - \intl{2'} \cmb(\veclb')
L(\veclb,\veclb')\right) \cmb(\veclc)  \cmb(\vecld) \Big> + {\rm Perm.}
\nonumber \\
&& \quad =  \Big<
\intl{1'} \cmb(\vecla') L(\vecla,\vecla') 
\intl{2'} \cmb(\veclb') L(\veclb,\veclb') \cmb(\veclc)  \cmb(\vecld) \Big> +
{\rm Perm.} \nonumber \\
\label{eqn:tri}
\end{eqnarray}
As written, to the lowest order, we find that contributions come from the first order term in $L$
given in equation~(\ref{eqn:lfactor}).
We further simplify to obtain
\begin{eqnarray}
 \left< \tilde \cmb(\bfl_1) \ldots
       \tilde \cmb(\bfl_4)\right>_c &=&  \Big<
\intl{1'} \cmb(\vecla')
\len(\vecla-\vecla') \, [(\vecla - \vecla') \cdot \vecla'] 
\intl{2'} \cmb(\veclb') 
\len(\veclb-\veclb') \, (\veclb - \veclb') \cdot \veclb'
\cmb(\veclc)  \cmb(\vecld) \Big> \,  \nonumber \\
&=&  C_{l_3}^\cmb C_{l_4}^\cmb \Big<
\len(\vecla+\veclc) \len(\veclb+\vecld) \Big>
\, [(\vecla + \veclc) \cdot \veclc] \, [(\veclb + \vecld) \cdot \vecld] + \, {\rm Perm.}\, , 
\end{eqnarray}
where there is an additional term through a 
permutation involving the interchange of
$\vecla+\veclc$ with $\vecla+\vecld$. Introducing the power spectrum of
 lensing potentials, we further simplify to 
obtain the CMB trispectrum due to gravitational lensing:
\begin{eqnarray}    
\tilde T^\cmb(\bfl_1,\bfl_2,\bfl_3,\bfl_4) = -C_{l_3}^{\cmb} C_{l_4}^{\cmb} \Big[ 
C^\pp_{|\vecl_1+\vecl_3|} (\vecla +\veclc) \cdot \veclc (\vecla + \veclc) \cdot \vecld  
 + C^\pp_{|\vecl_2+\vecl_3|} (\veclb +\veclc) \cdot \veclc (\veclb +\veclc)  \cdot \vecld \Big] 
+ \, {\rm Perm.} \, , 
\label{eqn:trilens}
\end{eqnarray}
where the permutations now contain 5 additional terms with the replacement of
$(l_3,l_4)$ pair by other combination of pairs.

The non-Gaussian contribution to the trispectrum, through
coupling of lensing deflection angle to secondary effects,
can be calculated with the replacement of, say, $\tilde \cmb(\veclc)$ and $\tilde \cmb(\vecld)$ in equation~(\ref{eqn:tri}) by $\cmb^\tot(\veclc)$ and $\cmb^\tot(\vecld)$ containing the sources of secondary fluctuations, which will
produce an addition contribution involving the $\cmb^s$ contribution associated with $\cmb^\tot$. Note that 
we can no longer consider cumulants such as $\langle \cmb(\vecla') \cmb^{\rm s}(\veclc)\rangle$
as the secondary effects are decoupled from recombination where primary fluctuations are imprinted. However,
contributions come from the correlation between $\cmb^{\rm s}$ and the lensing deflection $\phi$.
Here, contributions of equal importance come from both the first and second order terms in $L$ written in
equation~(\ref{eqn:lfactor}).   First, we note
\begin{eqnarray}
\left< \cmb^\tot(\bfl_1) \ldots
       \cmb^\tot(\bfl_4)\right>_c &=& \Big< \left( \cmb(\vecla) - \intl{1'} \cmb(\vecla')
L(\vecla,\vecla')\right) \left(\cmb(\veclb) - \intl{2'} \cmb(\veclb')
L(\veclb,\veclb')\right) \cmb^\s(\veclc)  \cmb^\s(\vecld) \Big>
+ {\rm Perm} \nonumber \\
&=& - C_{l_1}^\cmb  \left< L(\veclb,-\vecla)\cmb^s(\veclc) \cmb^\s(\vecld) \right> 
- C_{l_2}^\cmb \left< L(\vecla,-\veclb) \cmb^s(\veclc)  \cmb^\s(\vecld) \right>  \nonumber \\
&& \quad + \intl{1'} C_{l_1'}^\cmb \left< L(\veclb,-\vecla') 
L(\vecla,\vecla') \cmb^s(\veclc)  \cmb^\s(\vecld) \right> + {\rm Perm.}
\nonumber \\
\label{eqn:trisec}
\end{eqnarray}
Contributions to the trispectrum from the first two terms come
through the second order term in $L$, with the two $\phi$ terms
coupling to $\cmb^{\rm s}$. In the last term, contributions come
from the first order term of $L$ similar to the pure
 lensing contribution to the trispectrum.

After some straightforward simplifications, we write the
connected part of the trispectrum involving lensing-secondary coupling as
\begin{eqnarray}    
&& T^{\cmb \tot}(\bfl_1,\bfl_2,\bfl_3,\bfl_4) = \nonumber \\
&&  C_{l_3}^{\len\s} C_{l_4}^{\len\s} \Big\{  C^\cmb_{l_1} (\veclc \cdot \vecla) (\vecld \cdot \vecla) + 
 C^\cmb_{l_2} (\veclc \cdot \veclb) (\vecld \cdot \veclb) 
+ \left[ \veclc \cdot (\vecla + \veclc) \right]
\left[ \vecld \cdot (\veclb +\vecld) \right] C^\cmb_{|\vecl_1 + \vecl_3|} 
+\left[ \vecld \cdot (\vecla + \vecld) \right]
\left[ \veclc \cdot (\veclb +\veclc) \right] C^\cmb_{|\vecl_1 + \vecl_4|}
\Big\} \nonumber \\
&& \quad \quad + \, {\rm Perm.} \, 
\label{eqn:trisecfinal}
\end{eqnarray}
Note that the first two terms come from the first and 
second term in equation~(\ref{eqn:trisec}), 
while the last two terms in above are from the third term. 
As before, through permutations,
 there are five additional terms involving the pairings of $(l_3,l_4)$.

\begin{figure}[t]
\centerline{\psfig{file=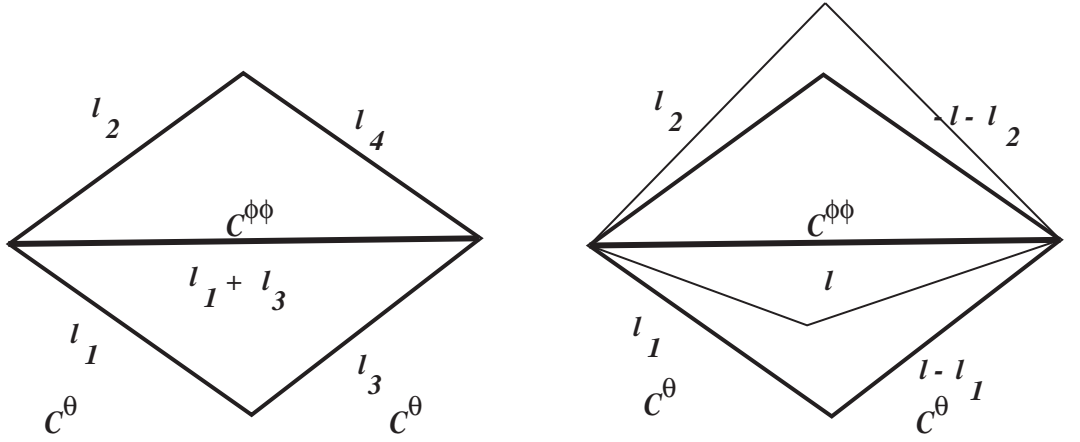,width=7.0in,angle=0}}
\caption{{\it Left:} The lensing trispectrum configuration, following
equation~(\ref{eqn:trilens}), where the diagonal of the configuration contains information related to
lensing while the sides
of the quadrilateral contain information related to primary anisotropy
power spectra at the last scattering surface. {\it Right:} This lensing information can be
extracted by a statistic that effectively probes the diagonal, such as the
power spectrum of squared temperatures discussed here. This statistic
sums over all quadrilateral configurations, as indicated
by thin lines, for a given length of the diagonal. As we find later, the
lensing information associated with the other --- vertical in this diagram --- diagonal,
proportional to $C_{|\vecl_1+\vecl_2|}^{\phi \phi}$, acts as a source of
non-Gaussian noise in this extraction. The other contribution to the non-Gaussian noise, through a term 
proportional to $C_{|\vecl-\vecl_1+\vecl_2|}$, comes from interchanging
$\vecl_1$ with $\vecl-\vecl_1$, which alters the vertical diagonal while
keeping the horizontal diagonal fixed.}
\label{fig:config}
\end{figure}

\section{Power spectrum of squared temperatures}
\label{sec:covariance}

As written in equation~(\ref{eqn:trilens}),  the trispectrum formed by lensing alone depends on 
the power spectrum of deflection angle.
Thus, with the right configuration for the CMB trispectrum, one can hope to extract some information related to lensing, mainly
$C_l^{\phi \phi}$, given some knowledge on CMB anisotropies, such as $C_l^{\cmb}$. Note that a contribution to the  CMB trispectrum due to
lensing comes from $C_l^{\phi \phi}$ when $l=|\vecl_i+\vecl_j|$ when $\vecl_i$ and $\vecl_j$ are the two sides of the 
quadrilateral that form the trispectrum. As illustrated in figure~(\ref{fig:config}), 
this is equivalent to the diagonal of the quadrilateral and can be probed easily with
a statistic that averages over the  configurations for a given length of the diagonal.
For this purpose, an useful option is to consider the configurations of the trispectrum that contribute to the
power spectrum of squared temperatures. To see why such a statistic probe the diagonal of the trispectrum,
 consider the definition of this squared power spectrum
\begin{equation}
\langle \tilde \cmb^2(\vecl) \tilde \cmb^2(\vecl') \rangle = (2\pi)^2
\delta_D(\vecl+\vecl') C_l^2\, ,
\end{equation}
where the Fourier transform of the squared temperature can be
written as a convolution of the temperature transforms
\begin{equation}
\tilde \cmb^2(\vecl) = \int \frac{d^2\vecl_1}{(2\pi)^2}
\tilde \cmb(\vecl_1)\tilde \cmb(\vecl-\vecl_1) \, .
\label{eqn:cmbsqr}
\end{equation}
Here, it should be understood that $\tilde \cmb^2(\vecl)$ refers to the
Fourier transform of the square of the temperature rather than square of the
Fourier transform of temperature, $[\tilde \cmb(\vecl)]^2$.
Note that the Fourier transform of
the temperature fluctuation, in the flat sky, is
\begin{equation}
\tilde \cmb(\vecl) = \int d^2\theta e^{-i\vecl \cdot \theta} \tilde T(\theta) \, .
\end{equation}

To compute the squared power spectrum of CMB temperature,
we take
\begin{eqnarray}
\langle \tilde \cmb^2(\vecl) \tilde \cmb^2(\vecl') \rangle &=& (2\pi)^2
\delta_D(\vecl+\vecl') C_l^2 \, \nonumber \\
&=& \int \frac{d^2\vecl_1}{(2\pi)^2}\int \frac{d^2\vecl_2}{(2\pi)^2}
\langle \tilde \cmb(\vecl_1) \tilde \cmb(\vecl-\vecl_1)
\tilde \cmb(\vecl_2)\tilde \cmb(\vecl'-\vecl_2) \rangle \, .\nonumber \\
\label{eqn:flat}
\end{eqnarray}
Since the lensing power spectrum contributes to the above squared power spectrum via the diagonals of the quadrilateral, 
the power spectrum of
squared temperatures contain terms which are simply proportional to $C_l^{\phi \phi}$ where $l$ is the multipole at which
the squared power spectrum is measured. As discussed in \cite{Hu01}, this then
allows an extraction of the angular power spectrum of lensing deflections.

There is also a dominant noise contribution to the squared power spectrum. This noise contribution comes from the
lowest order correlations and is present even in the presence of Gaussian temperature fluctuations; thus, we denote it  as
the Gaussian noise contribution. In the presence of the non-Gaussian contribution resulting from the lensing effect, 
we write these two contributions 
as
\begin{equation}
C_l^2 = C_l^{\rm G}+C_l^{\rm NG} \, ,
\end{equation}
with
\begin{eqnarray}
C_l^{2 \rm G} = \int \frac{d^2\vecl_1}{(2\pi)^2}  2C_{l_1}^\tot C_{|\vecl-\vecl_1|}^\tot \, ,
\end{eqnarray}
and the non-Gaussianity  captured by the four-point correlator as
\begin{eqnarray}
C_l^{2 \rm NG} = \int \frac{d^2\vecl_1}{(2\pi)^2}
\int \frac{d^2\vecl_2}{(2\pi)^2} \tilde T(\vecl_1,\vecl-\vecl_1,\vecl_2,-\vecl-\vecl_2) \, .
\end{eqnarray}.

Following our previous discussion on the lensing trispectrum, we can now write the full four-point contribution to the
squared power spectrum  --- by replacing $\vecl_1,\vecl_2,\vecl_3$ and $\vecl_4$ in equation~(\ref{eqn:trilens}) 
with the required configuration
from above --- such that
\begin{eqnarray}
&& C_l^{2 \rm NG} = 
C_l^{\phi\phi} \left[\intl{1} \vecl \cdot \vecl_1 C^\cmb_{l_1} + \vecl \cdot (\vecl-\vecl_1) C^\cmb_{|\vecl-\vecl_1|}\right]^2 \nonumber \\
&+& \intl{1} \intl{2} \Big\{ C_{|\vecl_1+\vecl_2|}^{\phi\phi}
[\vecl_1 \cdot (\vecl_1+\vecl_2) C^\cmb_{l_1} + \vecl_2 \cdot (\vecl_1+\vecl_2) C^\cmb_{l_2}] 
[(\vecl+\vecl_2) \cdot (\vecl_1+\vecl_2) C^\cmb_{|\vecl+\vecl_2|} - (\vecl_1+\vecl_2) \cdot (\vecl-\vecl_1) C^\cmb_{|\vecl-\vecl_1|}] \nonumber \\
&+& C_{|\vecl-\vecl_1+\vecl_2|}^{\phi\phi}
[(\vecl-\vecl_1 +\vecl_2) \cdot \vecl_2 C^\cmb_{l_2} + (\vecl-\vecl_1+\vecl_2) \cdot (\vecl-\vecl_1) C^\cmb_{|\vecl-\vecl_1|}]
 [-(\vecl-\vecl_1+\vecl_2) \cdot \vecl_1 C^\cmb_{l_1} + (\vecl-\vecl_1+\vecl_2) \cdot (\vecl+\vecl_2) C^\cmb_{|\vecl+\vecl_2|}] \Big\} \, . \nonumber \\
\end{eqnarray}

\begin{figure*}[t]
\centerline{\psfig{file=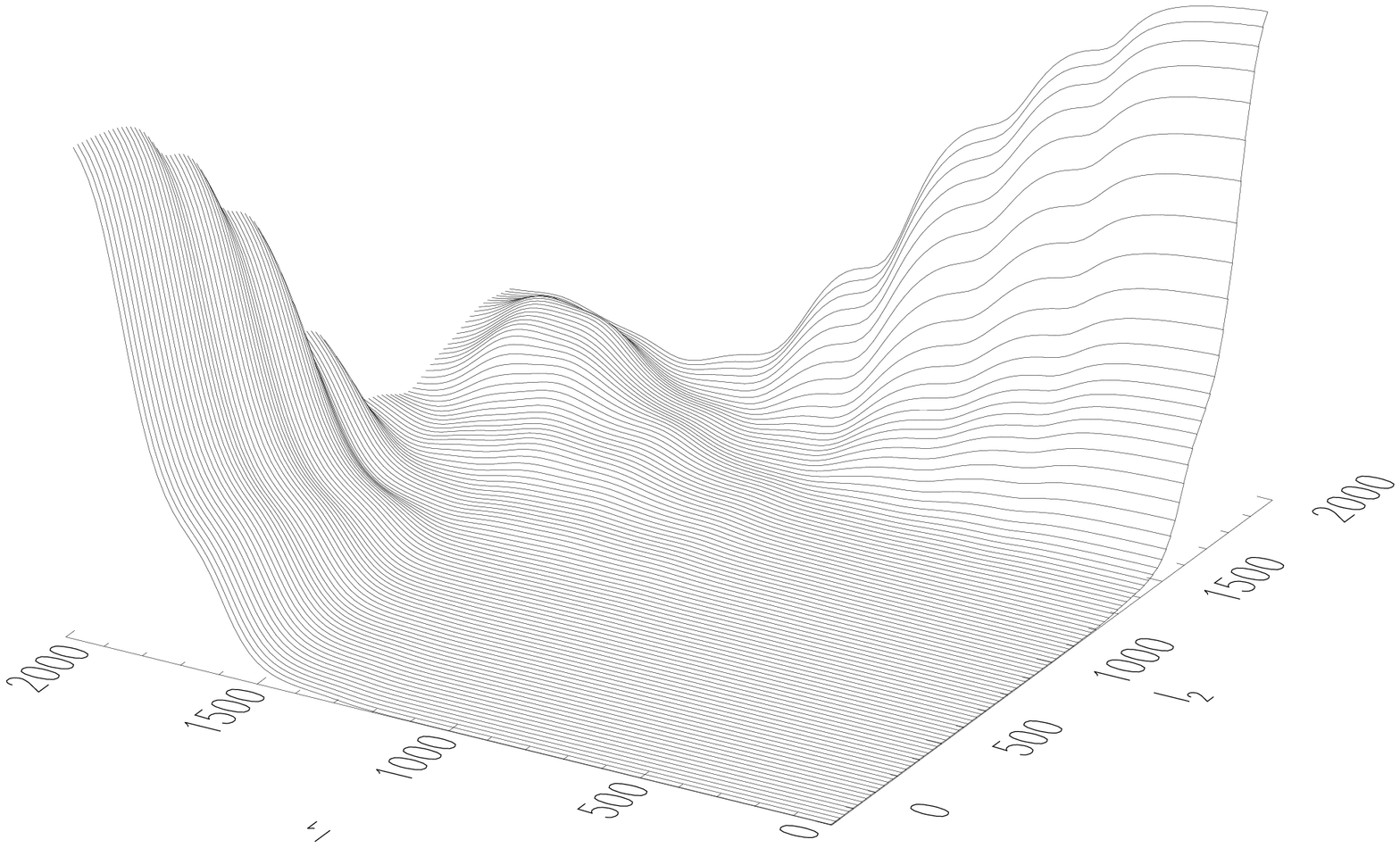,width=3.8in,angle=90}
\psfig{file=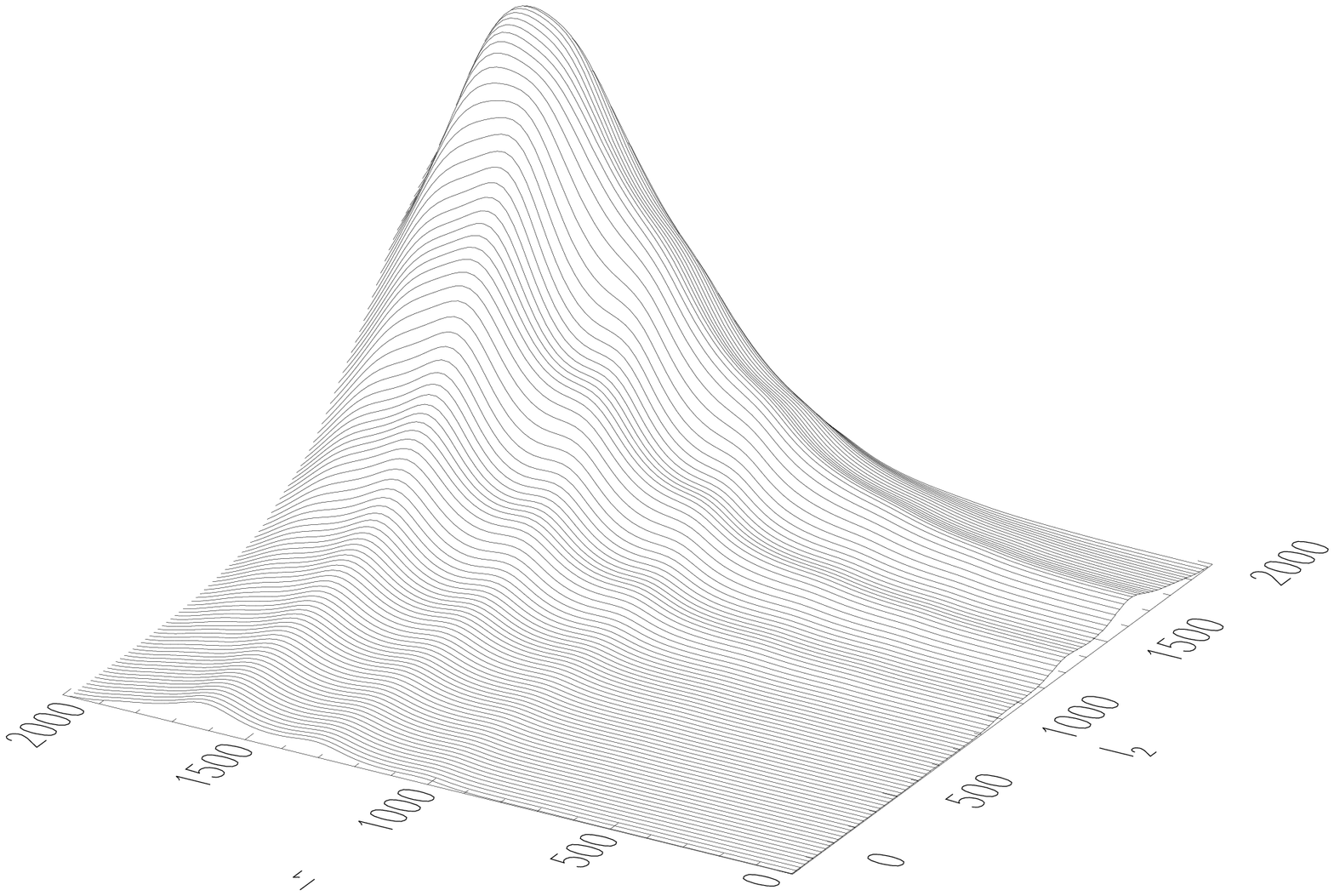,width=3.8in,angle=90}}
\caption{Optimal filter function $W(\vecl_1+\vecl_2, \vecl_1)$ for the
extraction of $C_l^{\phi\phi}$ from the CMB$^2$-CMB$^2$ power spectrum 
as a function of $l_1$ and $l_2$.  The angle between $\vecl_1$ and $\vecl_2$
is constrained by the fixed value of $l = |\vecl_1 + \vecl_2|$ in each plot.
The left plot $(l=100)$
and right plot $(l=2000)$ correspond to large and small angular scales
respectively. The optimal filter removes excess noise at
multipoles less than $\sim$ 1500 and is weighted to
exploit the sensitivity to lensing of the damping tail.}
\label{fig:filter}
\end{figure*}

As written, and discussed above, one of the terms in the non-Gaussian contribution to the 
squared power spectrum is directly proportional to $C_l^{\phi\phi}$. 
To extract this angular power spectrum of lensing deflections, consider the approach suggested by \cite{Hu01} involving the
quadratic statistics formed by the divergence of the temperature-weighted gradients. 
To understand the mechanism involved, consider a filtered version of the squared power spectrum such that
\begin{equation}
\hat{\cmb}^2(\vecl) = \int \frac{d^2\vecl_1}{(2\pi)^2}
W(\vecl, \vecl_1) \tilde \cmb(\vecl_1) \tilde \cmb(\vecl-\vecl_1) \, ,
\end{equation}
where $W$ is a filter that acts on the CMB temperature. This filtering can be either in real space or Fourier, or multipolar, space. For simplicity, we have introduced filters that are applicable in Fourier space.
To extract the optimal filter in the presence of the Gaussian noise alone,
we ignore the additional noise contribution from the second and third terms of the trispectrum involving integrals over
$C_{|\vecl_1+\vecl_2|}^{\phi \phi}$ and $C_{|\vecl-\vecl_1+\vecl_2|}^{\phi \phi}$.

We can, thus, write the filtered version of the squared power spectrum as
\begin{eqnarray}
\hat{C}_l^2 &=&  \int \frac{d^2\vecl_1}{(2\pi)^2} C_{l_1}^\tot
C_{|\vecl-\vecl_1|}^\tot W(\vecl, \vecl_1) \left[ W(-\vecl, -\vecl_1) +
W(-\vecl, \vecl_1 - \vecl)  \right] \nonumber \\
&+& C_l^{\phi\phi} \left[\int \frac{d^2\vecl_1}{(2\pi)^2}  W(\vecl, \vecl_1)
[\vecl \cdot \vecl_1 C^\cmb_{l_1} + \vecl \cdot (\vecl-\vecl_1) C^\cmb_{|\vecl-\vecl_1|}]\right] 
\left[\int \frac{d^2\vecl_1}{(2\pi)^2} W(-\vecl, -\vecl_1)
[\vecl \cdot \vecl_1 C^\cmb_{l_1} + \vecl \cdot (\vecl-\vecl_1) C^\cmb_{|\vecl-\vecl_1|}]\right]. \nonumber \\
\end{eqnarray}
We assume that our filter has the symmetry property
$W(\vecl, \vecl_1) = W(-\vecl, -\vecl_1) = W(-\vecl, \vecl_1 - \vecl)$ and then
check to see if this assumption is justified. To be consistent with prior results, 
we also make one more assumption here. We take the Gaussian noise
to be the dominant noise contribution and ignore the noise from the two extra terms in the non-Gaussian trispectrum
that involves integrals over 
$C^{\phi \phi}_{|\vecl_1+\vecl_2|}$ and $C^{\phi \phi}_{|\vecl-\vecl_1+\vecl_2|}$. 
Thus, the filter we will derive here is only optimal in the presence of the dominant Gaussian noise and
is sub-optimal when all noise terms are considered. We will later add in these two terms as an
extra source of noise, with the filter derived by ignoring them applied appropriately.

With these assumptions, we can write the signal-to-noise ratio for the detection of the $\phi$ power
spectrum per each multipole moment as,
\begin{eqnarray}
&& \left(\frac{{\rm S}}{{\rm N}}\right)^2 = \frac{(2l+1)}{2} \left( 
\frac{C_l^{\phi\phi} \left[\int \frac{d^2\vecl_1}{(2\pi)^2} W(\vecl, \vecl_1)
[\vecl \cdot \vecl_1 C^\cmb_{l_1} + \vecl \cdot (\vecl-\vecl_1) C^\cmb_{|\vecl-\vecl_1|}]\right]^2}{
C_l^{\phi\phi} \left[\int \frac{d^2\vecl_1}{(2\pi)^2} W(\vecl, \vecl_1)
[\vecl \cdot \vecl_1 C^\cmb_{l_1} + \vecl \cdot (\vecl-\vecl_1) C^\cmb_{|\vecl-\vecl_1|}]\right]^2+
 \int \frac{d^2\vecl_1}{(2\pi)^2} W(\vecl, \vecl_1)^2 2C_{l_1}^\tot C_{|\vecl-\vecl_1|}^\tot}\right)^2 \, .
\end{eqnarray}
The optimal filter in the presence of Gaussian noise is the one that will maximize this ratio. 

\begin{figure}[t]
\centerline{\psfig{file=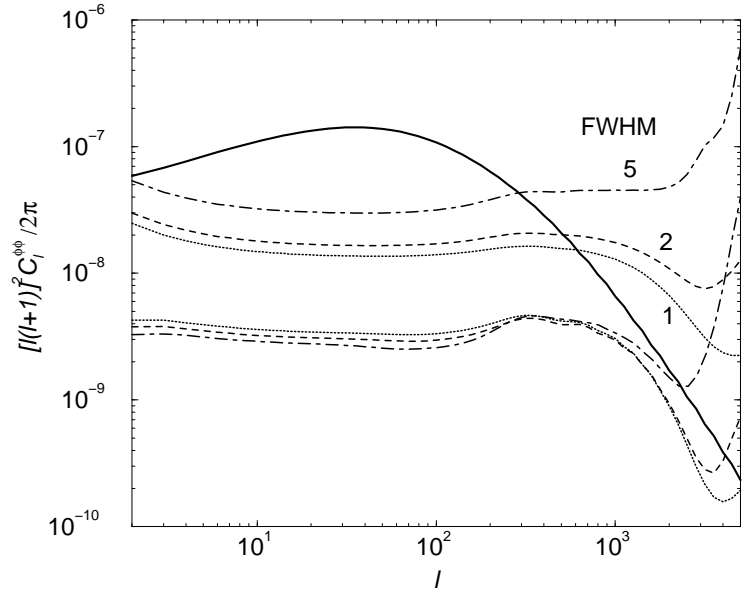,width=3.8in,angle=-90}}
\caption{The extraction of the lensing power spectrum from temperature data
alone. We show the
dominant Gaussian noise (top lines) and the noise associated with the extra
terms in the trispectrum due to lensing alone (bottom lines).
The curves are for three values of the resolution in the CMB temperature map
with an effective sensitivity of 1 $\mu$K $\sqrt{\rm sec}$.}
\label{fig:tempnoise}
\end{figure}

To extract the required filter, we take derivatives with respect to $W$ and find that the maximum is satisfied when
\begin{equation}
W(\vecl, \vecl_1) = \frac{[\vecl \cdot \vecl_1 C_{l_1}^\cmb + \vecl \cdot (\vecl-\vecl_1) C_{|\vecl-\vecl_1|}^\cmb]}{2C_{l_1}^\tot C_{|\vecl-\vecl_1|}^\tot}\,.
\end{equation}
Performing the explicit substitutions $\vecl \to -\vecl, \vecl_1 \to -\vecl_1$
and $\vecl \to -\vecl, \vecl_1 \to \vecl_1 - \vecl$, we see that our optimal
filter does possess the desired symmetry.
We introduced this filter in Ref. \cite{Coo01} when discussing the CMB$^2$-SZ power spectrum as  a way to 
extract the lensing-SZ correlation directly using the power spectrum of squared temperature and temperature. 
It is not a surprise that the same filter appears now when extracting the lensing-lensing correlations using the
power spectrum of squared temperature and squared temperature. 
A physical mechanism to achieve this filter in real space is presented in Ref. \cite{Hu01b} by taking the
gradient of the temperature, weighting it by a noise filtered CMB map, then taking the divergence of this product.

With the filter applied, we get a direct estimator of $C_l^{\phi\phi}$. Introducing,
\begin{eqnarray}
&& N_l^{-1} \equiv \int \frac{d^2\vecl_1}{(2\pi)^2}  
\frac{
[\vecl \cdot \vecl_1 C^\cmb_{l_1} + \vecl \cdot (\vecl-\vecl_1) C^\cmb_{|\vecl-\vecl_1|}]^2}
{2C_{l_1}^\tot C_{|\vecl-\vecl_1|}^\tot} \nonumber \\
\label{eqn:nl}
\end{eqnarray}
The filtered squared power spectrum is such that
\begin{equation}
N_l^2 \hat{C}_l^2 =  N_l + C_l^{\phi\phi} \, ,
\end{equation}
where the Gaussian noise associated with the extracted $C_l^{\phi\phi}$ is $N_l$.

We can now add in the noise contribution arising from the two extra terms in the non-Gaussian trispectrum
which we dropped when deriving the optimal filter in the presence of the Gaussian noise only.
This extra noise contribution, neglected in prior studies, is
\begin{eqnarray} \label{E:FNG}
&& N_l^{\rm NG} =  N_l^2 \int \frac{d^2\vecl_1}{(2\pi)^2} \int \frac{d^2\vecl_2}{(2\pi)^2} W(\vecl, \vecl_1)W(-\vecl, \vecl_2)\nonumber \\
&&\Big\{ C_{|\vecl_1+\vecl_2|}^{\phi\phi}
[\vecl_1 \cdot (\vecl_1+\vecl_2) C^\cmb_{l_1} + \vecl_2 \cdot (\vecl_1+\vecl_2) C^\cmb_{l_2}]
[(\vecl+\vecl_2) \cdot (\vecl_1+\vecl_2) C^\cmb_{|\vecl+\vecl_2|} - (\vecl_1+\vecl_2) \cdot (\vecl-\vecl_1) C^\cmb_{|\vecl-\vecl_1|}] \nonumber \\
&+& C_{|\vecl-\vecl_1+\vecl_2|}^{\phi\phi}
[(\vecl-\vecl_1 +\vecl_2) \cdot \vecl_2 C^\cmb_{l_2} + (\vecl-\vecl_1+\vecl_2) \cdot (\vecl-\vecl_1) C^\cmb_{|\vecl-\vecl_1|}]
[-(\vecl-\vecl_1+\vecl_2) \cdot \vecl_1 C^\cmb_{l_1} + (\vecl-\vecl_1+\vecl_2) \cdot (\vecl+\vecl_2) C^\cmb_{|\vecl+\vecl_2|}] \Big\} \, . \nonumber \\
\end{eqnarray}

With the non-Gaussian noise, we can write
\begin{equation}
N_l^2 \hat{C}_l^2 =  N_l + N_l^{\rm NG}+ C_l^{\phi\phi} \, .
\end{equation}
There is a strong possibility that the noise-bias $N_l$ associated with the
Gaussian contribution to the squared power spectrum can be removed during
the lensing reconstruction. This would involve the construction of
a new estimator of lensing, or filtering scheme, that removes $N_l$ and
leaves $N_l^{\rm NG}$ as the final noise
contribution associated with the reconstructed $C_l^{\phi\phi}$.
The removal of the dominant Gaussian noise bias is extremely useful since this
allows one to
lower the noise contribution associated with $C_l^{\phi \phi}$ by at least an
order of magnitude. We will return later to
the impact of such a noise-bias removal on the identification of an
inflationary gravitational wave signature in the CMB in the form
of a curl (or B-mode) contribution to the polarization.

In addition to the additional non-Gaussian noise contribution from the lensing
trispectrum itself, there is an additional noise contribution
resulting from the relevant configuration of the trispectrum due to
lensing-secondary correlations. 
We do not reproduce this contribution here, but it can be easily calculated by
replacing $(\vecl_1,\vecl_2,\vecl_3,\vecl_4)$ in
equation~(\ref{eqn:trisecfinal}), with
$(\vecl_1,\vecl-\vecl_1,\vecl_2,-\vecl-\vecl_2)$.
Note that this additional noise results from the correlation between potentials
that deflect CMB photons and the
secondary contributions associated with the ISW and SZ thermal effects. 
In the presence of such contributions, we also note that $C_l^\tot$ in
equation~(\ref{eqn:nl}), contains all contributions to noise,
including the secondary contribution. It is well known that certain secondary
contributions can be removed based on
distinct signatures, such as the spectral dependence of the SZ effect
\cite{Cooetal00}. In such cases, we expect only
the lensing-ISW correlation to produce a noise contribution. This effect is
small and can be easily ignored.
In the presence of the SZ contribution, however, there is a significant noise
contribution that can potentially limit the
extraction of the lensing power spectrum.

Note that the filter is symmetric in $\vecl_1$ and $\vecl-\vecl_1$. We show surfaces of this filter in figure~(\ref{fig:filter})
as a function of $l_1$ and $l_2 =|\vecl-\vecl_1|$ for two values of $l$ corresponding to large and small angular scales
respectively. As shown, the filter behaves such that information related to lensing is extracted at arcminute scales
via the damping tail of the CMB anisotropies. Additionally, since lensing smooths the acoustic peak structure, further information is also extracted in a subtle way from fluctuations associated with large angular scales.  
Thus, a simple filter which effectively cuts off the acoustic peak structure
but extracts information from small angular scales, such a a step function in multipolar space, does poorly when
compared to the filtering scheme proposed above \cite{Coo01}. As mentioned before deriving the filter, it should be
noted that the proposed filter is only optimal to the
extent that Gaussian noise contribution is dominant. In the presence of noise
contributions
from additional four-point correlations or, more importantly, additional
terms in the lensing trispectrum itself, the filter is sub-optimal. We failed to
find a simple analytical description for an optimal filter that also suppresses
the
non-Gaussian noise contribution from the additional terms in the lensing
trispectrum.  As we find later, since these terms
only contribute at an order of magnitude below the dominant Gaussian noise,
we do not expect this to be a significant
problem in the extraction of the lensing information.

\section{The EB Trispectrum}
\label{sec:EBtri}

In addition to temperature, lensing also affects the CMB polarization. We
discuss here the best estimator from polarization 
observables for extracting the lensing contribution following the discussion
in Ref. \cite{HuOka01} and provide relevant
equations for other estimators in the Appendix.
The statistic considered in this section involves the squared map constructed by convolving, in Fourier
space, the E- and B-mode maps of the polarization. We denote this by
$(EB)(\vecl)$.
To understand this combination, we consider the Fourier space analogue  and, as before, write,
\begin{equation}
EB(\vecl) = \frac{1}{2}\int \frac{d^2\vecl_1}{(2\pi)^2} \left[E(\vecl_1) B(\vecl-\vecl_1) +
B(\vecl_1) E(\vecl-\vecl_1) \right] \, .
\end{equation}
Now considering the power spectrum formed by $\langle EB(\vecl) EB(\vecl')\rangle=(2\pi)^2 \delta(\vecl+\vecl') C_l^{(EB)^2}$,
we write
\begin{eqnarray}
\langle EB(\vecl) EB(\vecl')\rangle = \frac{1}{4} 
 \int \frac{d^2\vecl_1}{(2\pi)^2} \int \frac{d^2\vecl_2}{(2\pi)^2} 
\langle \left[E(\vecl_1) B(\vecl-\vecl_1) +B(\vecl_1) E(\vecl-\vecl_1) \right] 
\left[E(\vecl_2) B(\vecl'-\vecl_2) +B(\vecl_2) E(\vecl'-\vecl_2) \right] \rangle \, .
\end{eqnarray}
Here, again, we find two sources of contributions, involving a noise contribution due to Gaussian fluctuations in the E- and B-maps,
\begin{equation}
C_l^{(EB)^2 \rm G}
 = \frac{1}{4}\int\frac{d^2\vecl_1}{(2\pi)^2} \left[2C_{l_1}^{EE \tot}C_{|\vecl-\vecl_1|}^{BB \tot}+2C_{l_1}^{BB \tot}C_{|\vecl-\vecl_1|}^{EE \tot}\right] \, ,
\end{equation}
and a non-Gaussian contribution due to the lensing effect on polarization.

\begin{figure}[t]
\centerline{\psfig{file=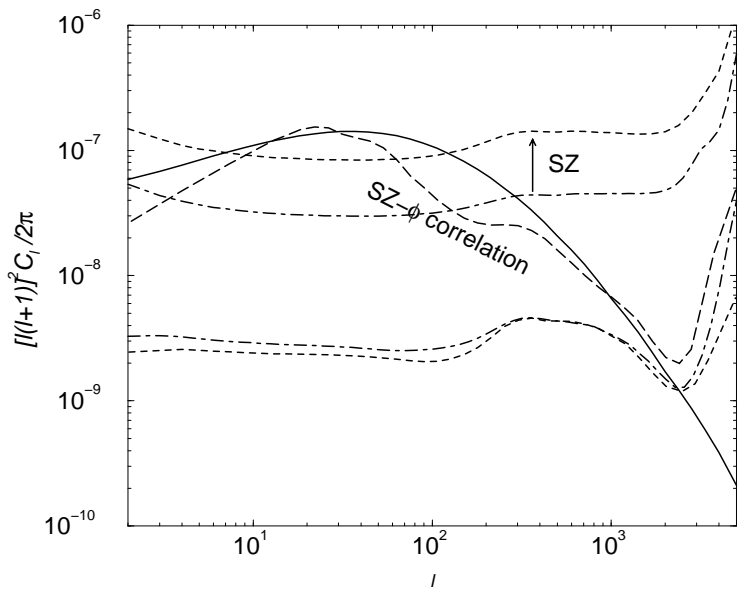,width=3.8in,angle=-90}}
\caption{The extraction of lensing power spectrum from temperature data in the presence of foregrounds (mainly the thermal
SZ effect). The increase in dominant Gaussian noise (top lines) is due to the extra SZ power spectrum, which peaks at
small angular scales. The dot-dashed line is the resulting non-Gaussian noise contribution due to the trispectrum formed
by the SZ-lensing cross-correlation.}
\label{fig:sznoise}
\end{figure}

\begin{figure}[t]
\centerline{\psfig{file=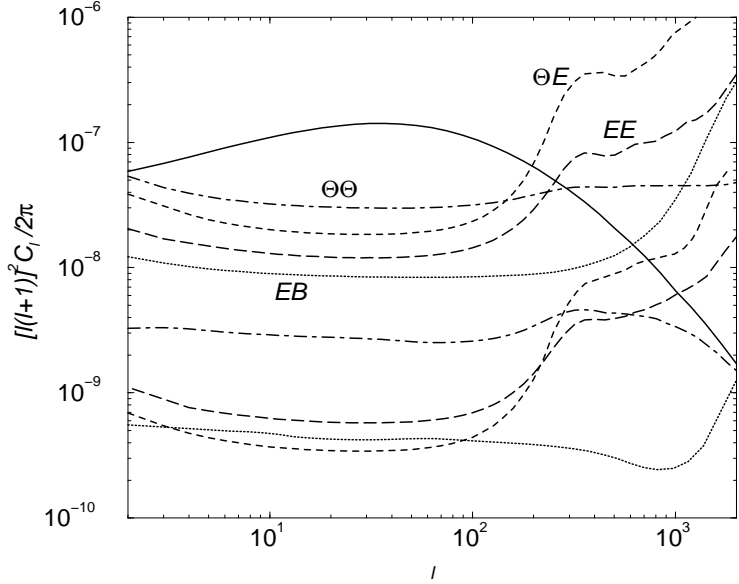,width=3.8in,angle=-90}}
\caption{The extraction of lensing power spectrum from polarization data. We show the
associated noise due to the dominant Gaussian noise (top lines) and noise due to extra terms in the trispectrum 
(bottom lines). The curves are for a resolution of 5 arcmin in the CMB temperature map with an effective sensitivity of 1 $\mu$K 
$\sqrt{\rm sec}$. As shown, the EB estimator probes the lensing potential power spectrum to smaller angular scales than the
quadratic estimator on temperature data alone or estimators based on other combinations of polarization and the temperature.}
\label{fig:ebnoise}
\end{figure}

To calculate the lensing contribution to the polarization, we follow the notation in Ref. \cite{Hu00}. Introducing,
\begin{eqnarray}
{}_{\pm} \tilde X(\bn) & = &  {}_{\pm}X(\bn + \nabla\len) \\
\label{eqn:ebl}
        & \approx &
{}_{\pm} X(\bn) + \nabla_i \len(\bn) \nabla^i {}_{\pm} X(\bn) + {1 \over 2} \nabla_i \len(\bn) \nabla_j \len(\bn) \nabla^{i}\nabla^{j}
{}_{\pm}  X(\bn) \,,
\nonumber
\end{eqnarray}
where $\spin{\pm} X = Q\pm i U$ represent combinations of the Stokes parameters. The lensing effect is again to move the photon 
propagation directions on the sky.   In Fourier space, we can consider the E- and B-mode decomposition introduced in Ref. \cite{KamKosSte97}
such that
${}_\pm X(\l) = E(\l)\pm i B(\l)$ and
\begin{eqnarray}
{}_\pm \tilde X(\l) &=& {}_\pm X(\l)
-
\intl{1}
{}_\pm X(\la)
 e^{\pm 2i (\varphi_{\vecl_1}- \varphi_{\vecl})} L(\l,\l_1) \,.
\end{eqnarray}
With this, we can write
\begin{eqnarray}
\tilde E(\vecl) &=& E(\vecl) - \intl{1} \left[E(\vecl_1) \cos 2 (\varphi_{\vecl_1} -\varphi_\vecl) - B(\vecl_1) \sin 2 (\varphi_{\vecl_1} -\varphi_\vecl)\right] \phi(\vecl-\vecl_1) (\vecl-\vecl_1) \cdot \vecl_1 \nonumber \\
\tilde B(\vecl) &=& B(\vecl) - \intl{1} \left[E(\vecl_1) \sin 2 (\varphi_{\vecl_1} -\varphi_\vecl) + B(\vecl_1) \cos 2 (\varphi_{\vecl_1} -\varphi_\vecl)\right] \phi(\vecl-\vecl_1) (\vecl-\vecl_1) \cdot \vecl_1 \, ,
\end{eqnarray}
Since primordial $B(\vecl)$, due to the gravitational wave contribution, is small, we can make the useful approximation that
\begin{eqnarray}
\tilde E(\vecl) &=& E(\vecl) - \intl{1} \left[E(\vecl_1) \cos 2 (\varphi_{\vecl_1} -\varphi_\vecl)\right] \phi(\vecl-\vecl_1) (\vecl-\vecl_1) \cdot \vecl_1
\nonumber \\
\tilde B(\vecl) &=& - \intl{1} \left[E(\vecl_1) \sin 2 (\varphi_{\vecl_1} -\varphi_\vecl)\right] \phi(\vecl-\vecl_1) (\vecl-\vecl_1) \cdot \vecl_1 \, ,
\end{eqnarray}
Under this approximation, the lensed polarization power spectra can be
expressed in terms of $C_l^{\phi\phi}$ and the unlensed quantities following Ref.
\cite{Hu00}:
\begin{eqnarray}
\tilde C_l^{EE} &=& \left[ 1 - \intl{1}
C^{\phi\phi}_{l_1} \left(\vecl_1 \cdot \vecl\right)^2 \right]	\, 
 	                        C_l^{EE}
        + \frac{1}{2} \intl{1} C_{|\vecl - \vecl_1|}^{\phi\phi}
[(\vecl - \vecl_1)\cdot \vecl_1]^2 
C^{EE}_{l_1} [1 + \cos(4\varphi_{\vecl_1})] \, ,\nonumber \\
\tilde C_l^{BB} &=& 
        \frac{1}{2} \intl{1} C_{|\vecl - \vecl_1|}^{\phi\phi}
[(\vecl - \vecl_1)\cdot \vecl_1]^2
C^{EE}_{l_1} [1 - \cos(4\varphi_{\vecl_1})]  \, .\nonumber \\
\end{eqnarray}
The total polarization power spectra is the sum of these lensed spectra and a
noise contribution
\begin{equation}
C_l^{EE\tot} = \tilde C_l^{EE} + C_l^\n \, \quad \quad C_l^{BB\tot} = \tilde C_l^{BB} + C_l^\n ,
\end{equation}
where the noise component is given by equation (\ref{E:noise}) with a possibly
different NET $s$ for polarization.

With the lensing effect on $B(\vecl)$, we can now calculate the trispectrum contribution to the squared power spectrum as
\begin{eqnarray}
&& C_l^{(EB)^2 \rm NG} = -\frac{1}{4}C_l^{\phi\phi} \left[\intl{1} \Big\{
C_{l_1}^{EE} \vecl \cdot \vecl_1 
+C_{|\vecl-\vecl_1|}^{EE} \vecl \cdot (\vecl-\vecl_1) \Big\} \sin 2 (\varphi_{\vecl_1}+ \varphi_{\vecl-\vecl_1}) \right]^2
\nonumber \\
&+&\intl{1} \intl{2} \Big\{\frac{1}{4}C_{|\vecl_1+\vecl_2|}^{\phi\phi} \left[C_{l_1}^{EE} C_{|\vecl+\vecl_2|}^{EE} (\vecl_1+\vecl_2) 
\cdot (\vecl+\vecl_2) (\vecl_1+\vecl_2) \cdot \vecl_1  - C_{l_2}^{EE} C_{|\vecl-\vecl_1|}^{EE} (\vecl_1+\vecl_2) \cdot \vecl_2 
(\vecl_1+\vecl_2) \cdot (\vecl-\vecl_1) \right] \nonumber \\
&& \quad \quad \times \sin 2(\varphi_{\vecl-\vecl_1} -\varphi_{\vecl+\vecl_2})  \sin 2(\varphi_{\vecl_1} +\varphi_{\vecl_2}) \nonumber \\
&+&\frac{1}{4}C_{|\vecl-\vecl_1+\vecl_2|}^{\phi\phi} \left[C_{|\vecl-\vecl_1|}^{EE} C_{|\vecl+\vecl_2|}^{EE} 
(\vecl-\vecl_1+\vecl_2) \cdot (\vecl+\vecl_2) (\vecl-\vecl_1+\vecl_2) \cdot (\vecl-\vecl_1)  - C_{l_1}^{EE} C_{l_2}^{EE} (\vecl-\vecl_1+\vecl_2) \cdot \vecl_1 
(\vecl-\vecl_1+\vecl_2) \cdot \vecl_2 \right] \nonumber \\
&& \quad \quad \times \sin 2(\varphi_{\vecl_1} -\varphi_{\vecl+\vecl_2})  \sin 2(\varphi_{\vecl-\vecl_1} +\varphi_{\vecl_2}) \Big\}\, ,
\end{eqnarray}

To extract the lensing contribution, one can again consider a filtering scheme, and following arguments similar to the case with temperature,
the derived filter that maximizes the signal-to-noise in the presence of Gaussian noise is
\begin{equation}
W(\vecl, \vecl_1) = \frac{\left[C_{l_1}^{EE} \vecl \cdot \vecl_1 
+C_{|\vecl-\vecl_1|}^{EE} \vecl \cdot (\vecl-\vecl_1)\right]\sin 2 (\varphi_{\vecl_1} +\varphi_{\vecl-\vecl_1})}{
C_{l_1}^{EE \tot} C_{|\vecl-\vecl_1|}^{BB \tot} + C_{l_1}^{BB \tot} C_{|\vecl-\vecl_1|}^{EE \tot} } \, .
\end{equation}
With this filter applied, the associated noise due to the dominant Gaussian part is $N_l$, where,
\begin{eqnarray}
N_l^{-1} = \intl{1} \frac{\left[C_{l_1}^{EE} \vecl \cdot \vecl_1 
+C_{|\vecl-\vecl_1|}^{EE} \vecl \cdot (\vecl-\vecl_1)\right]^2 \sin^2 2 (\varphi_{\vecl_1} + \varphi_{\vecl-\vecl_1})}{
2C_{l_1}^{EE \tot} C_{|\vecl-\vecl_1|}^{BB \tot} + 2C_{l_1}^{BB \tot} C_{|\vecl-\vecl_1|}^{EE \tot} } \, .
\end{eqnarray}
The extra noise due to the additional terms in the trispectrum is
\begin{eqnarray}
N_l^{\rm NG} &=& N_l^2 
 	\intl{1} \intl{2} W(\vecl, \vecl_1) W(-\vecl, \vecl_2) \Big\{ \nonumber \\
&\times& \frac{1}{4}C_{|\vecl_1+\vecl_2|}^{\phi\phi} \left[C_{l_1}^{EE} C_{|\vecl+\vecl_2|}^{EE} (\vecl_1+\vecl_2) 
\cdot (\vecl+\vecl_2) (\vecl_1+\vecl_2) \cdot \vecl_1  - C_{l_2}^{EE} C_{|\vecl-\vecl_1|}^{EE} (\vecl_1+\vecl_2) \cdot \vecl_2 
(\vecl_1+\vecl_2) \cdot (\vecl-\vecl_1) \right] \nonumber \\
&& \quad \quad \times \sin 2(\varphi_{\vecl-\vecl_1} -\varphi_{\vecl+\vecl_2})  \sin 2(\varphi_{\vecl_1} +\varphi_{\vecl_2}) \nonumber \\
&+&\frac{1}{4}C_{|\vecl-\vecl_1+\vecl_2|}^{\phi\phi} \left[C_{|\vecl-\vecl_1|}^{EE} C_{|\vecl+\vecl_2|}^{EE} 
(\vecl-\vecl_1+\vecl_2) \cdot (\vecl+\vecl_2) (\vecl-\vecl_1+\vecl_2) \cdot (\vecl-\vecl_1)  - C_{l_1}^{EE} C_{l_2}^{EE} (\vecl-\vecl_1+\vecl_2) \cdot \vecl_1 
(\vecl-\vecl_1+\vecl_2) \cdot \vecl_2 \right] \nonumber \\
&& \quad \quad \times \sin 2(\varphi_{\vecl_1} -\varphi_{\vecl+\vecl_2})  \sin 2(\varphi_{\vecl-\vecl_1} +\varphi_{\vecl_2})\Big\} \, .
\end{eqnarray}

\begin{figure}[t]
\centerline{\psfig{file=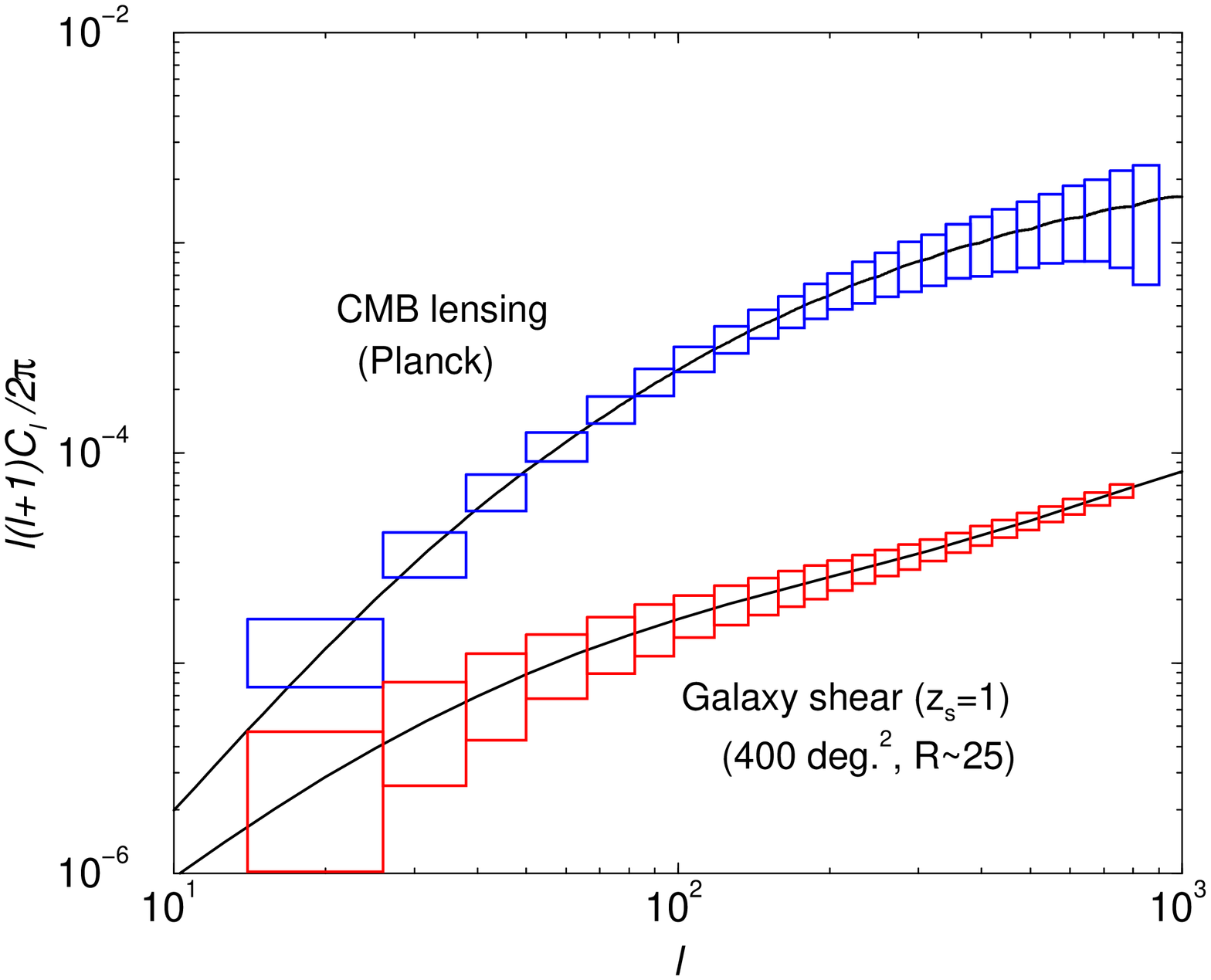,width=3.8in,angle=-90}
\psfig{file=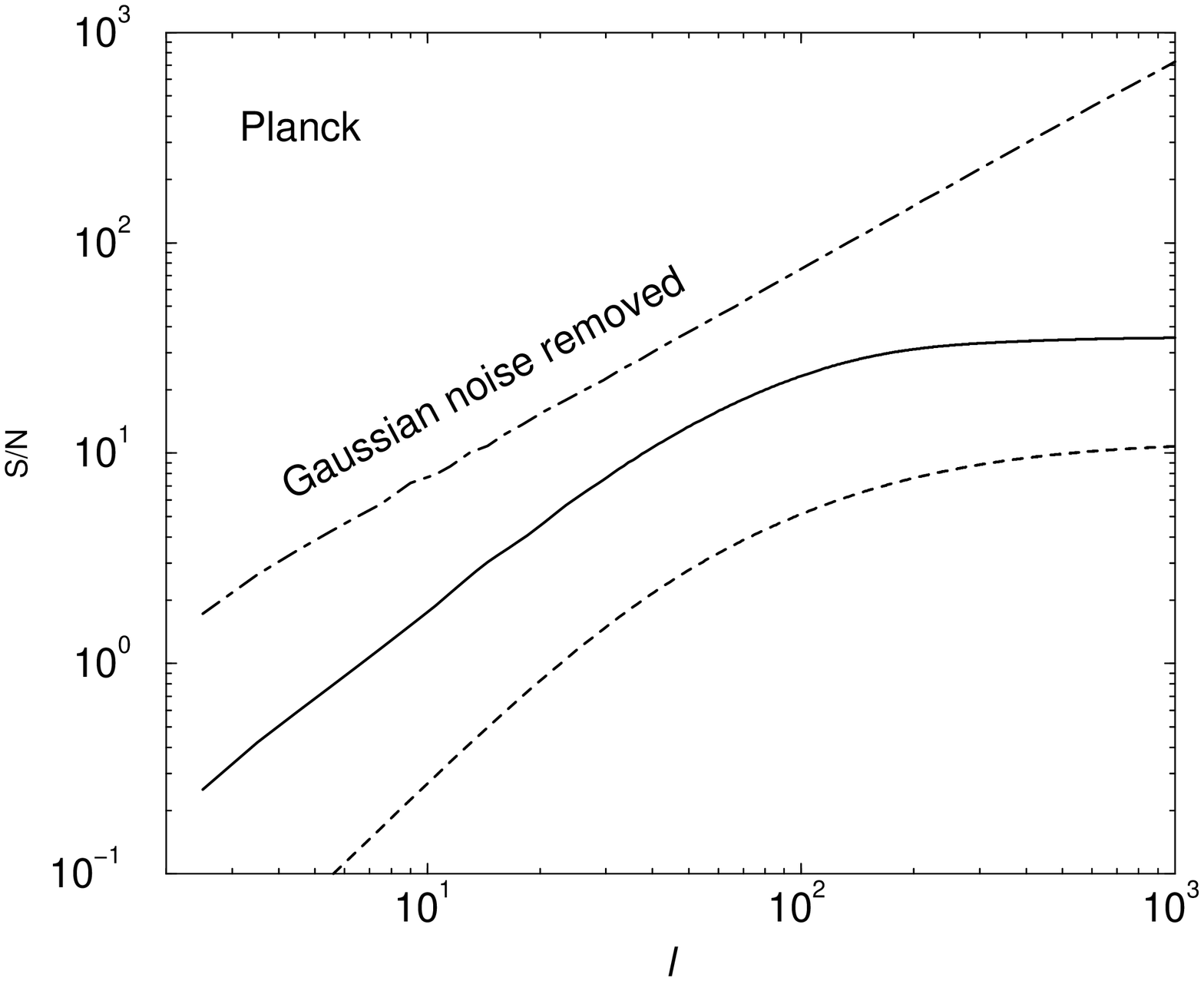,width=3.8in,angle=-90}}
\caption{{\it Left:} Weak lensing as a CMB experiment. We show errors for the reconstruction of convergence, or projected mass
density, to the last scattering surface with Planck temperature data. This is compared to the convergence measurements
expected from observations of galaxy ellipticities which probe to redshifts of $\sim$ 1 to 2. {\it Right:} The cumulative signal-to-noise ratio for the detection of the lensing effect in Planck data, using the filtered squared temperature statistic discussed here (solid line)
and the gradient statistic (dotted line) described in Ref. [14]
. The dot-dashed line is the cumulative signal-to-noise ratio when the dominant Gaussian noise
is removed as part of the lensing reconstruction. With such a removal, the noise in the lensing reconstruction is determined by the
non-Gaussian contributions which are roughly an order of magnitude below the Gaussian noise level.}
\label{fig:convergence}
\end{figure}

\section{Discussion}
\label{sec:results}

We have introduced the trispectrum, or four-point correlation function, of CMB anisotropies as a probe of the large scale structure
weak lensing effect. The trispectrum in CMB anisotropies is generated by the non-linear mode coupling behavior of the weak lensing
effect on the CMB. As written in equation (\ref{eqn:trilens}), the trispectrum
formed by lensing contains a significant contribution
from $C_l^{\phi \phi}$ when $l=|\vecl_1+\vecl_3|$.  $\vecl_1$ and $\vecl_3$
are the two sides of the quadrilateral that formed the
trispectrum in the Fourier space as illustrated in figure \ref{fig:config}.
Contributions thus probe the lensing effect
via the diagonal of the trispectrum configuration, and statistics sensitive to
the diagonal will provide a practical approach for extracting the lensing angular power spectrum.

A useful statistic for this purpose is the
power spectrum formed by squared temperatures introduced earlier.
This approach expands the 
suggestion in Ref. \cite{Coo01} to use the squared temperature-temperature power spectrum to study the CMB bispectrum and a filtered
version to extract the power spectrum associated with lensing-SZ correlation. 
The power spectrum formed by squared temperatures probes the trispectrum. This power spectrum, however, contains a dominant contribution
due to the  Gaussian fluctuations associated with temperature which potentially
dominates the contribution arising from the non-Gaussian
weak lensing effect. To optimally extract the lensing contribution in the
presence of this dominant
Gaussian noise, we have introduced a filter in Fourier space. This filter requires knowledge of the noise contributions to the map
as well as the power spectra of prelensed primary anisotropies at the last scattering surface. Though the filter can be
easily applied to the Fourier transform of the squared temperature map in the Fourier space, a useful direct approach is
presented in Ref. \cite{Hu01b}. This real space construction involves the divergence of the temperature-weighted 
temperature gradients, with each of these temperature estimators first filtered
for the excess noise. 

We have extended the previous discussions of the extraction of lensing signal
to include a discussion of the full trispectrum.
The full trispectrum contains terms which were previously ignored and these terms contribute as additional sources of
noise in the lensing reconstruction and can potentially interfere with the
extraction of the lensing signal from CMB data. These additional non-Gaussian noise terms correspond to the lensing contribution
associated with the second diagonal of the trispectrum configuration shown in figure~(\ref{fig:config}) and the
same diagonal under a permutation, while keeping the same vector $\vecl$ for the diagonal from which lensing information 
is extracted. In figure~(\ref{fig:tempnoise}), we summarize our results
related to the lensing extraction using temperature data alone. We show the dominant Gaussian noise contribution and the
associated non-Gaussian noise contribution, due to the two additional terms in the trispectrum which involve integrals over
$C^{\phi \phi}_{|\vecl_1+\vecl_2|}$ and $C^{\phi \phi}_{|\vecl-\vecl_1+\vecl_2|}$. We show these noise contributions for
temperature maps constructed with an effective detector sensitivity of 1 $\mu$K $\sqrt{\rm sec}$ and varying resolution
with beam widths of 5 to 1 arcmin. As shown, non-Gaussian noise contributions from additional terms in the
trispectrum are roughly an order of magnitude below the dominant Gaussian noise contribution. 

This conclusion is subject to several conditions. As summarized in figure~(\ref{fig:sznoise}), the extraction from
temperature alone can be significantly impaired by the presence of foregrounds
that correlate with the lensing deflection angle.
For example, the SZ contribution hinders the separation in two ways.
The first is an enhancement of the Gaussian noise signal by the addition of
$C_l^{\sz}$ to $C_l^\tot$ as indicated by the
difference between top dashed and the dot-dashed lines.
The second is a noise contribution due to the non-Gaussian trispectrum formed
by the correlation of lensing potentials with the SZ effect due to the large
scale structure. 
We show this with the long-dashed line. In the presence of SZ, it is now clear that the lensing extraction is strongly
limited. The other source of correlation is the ISW effect. Its effects, through the contribution to Gaussian
noise, are already included in the $C_l^\cmb$ defined earlier.  The power
spectra calculated using the publicly available CMBFAST code \cite{SelZal96}
 contain the ISW signal as well. The noise due to the correlation between
ISW and lensing is significantly smaller  and is at a level below $10^{-10}$ in figure~(\ref{fig:sznoise}). Thus, one can
easily ignore the additional noise contributions resulting from the ISW effect.
Furthermore, the SZ noise contributions can also be controlled. One can easily separate the SZ contribution from
thermal CMB fluctuations based on the frequency spectrum of the SZ effect \cite{Cooetal00}. In the upcoming
Planck mission, SZ separation based on multifrequency data will prevent this contribution from adding significant noise to CMB maps.
We suggest that small angular scales experiments seeking to
extract the lensing signal based on temperature anisotropies be designed to
contain channels that are optimized for a separation of the SZ signal from thermal CMB fluctuations.

In addition to temperature, lensing also modifies polarization data. Thus, one can employ estimators based on polarization, and
combinations of polarization and temperature, to extract information related to lensing. 
In figure~(\ref{fig:ebnoise}), we illustrate the usefulness of polarization observations for a lensing separation.
The dotted lines are the errors associated with the extraction based on the EB estimator.
This estimator was suggested by
\cite{HuOka01} as an estimator that probes the angular power spectrum of
lensing on scales smaller than that probed by
temperature or other combinations of temperature and polarization.
EE and $\cmb$E estimators also probe the lensing deflection power spectrum, and
have Gaussian and non-Gaussian noise contribution in between those of the
$\cmb \cmb$ and EB estimators.
While we presented a detailed derivation of the EB estimator,
we write down  the trispectrum and related filters for other estimators associated with polarization and combinations of 
polarization and temperature in the Appendix.
As shown in figure~(\ref{fig:ebnoise}), 
we again find the non-Gaussian noise contributions due to additional noise terms in various trispectra due to lensing alone to be 
roughly an order of magnitude below the dominant Gaussian noise contribution. 

Similar to the $\cmb \cmb$ estimator, 
the $\cmb$E estimator contains an additional noise term due to the correlation between weak lensing and secondary anisotropies
such as the SZ effect. We  have presented a derivation of this contribution in
the Appendix.
In both these cases, note that we have only considered the secondary contributions to the temperature.
Unlike contributions to the temperature,
 secondary contributions to the polarization are significantly smaller and are higher order in density fluctuations;
for example, in addition to the density field, these secondary effects also depend on the velocity field \cite{Hu00b}.
This makes the secondary polarization-lensing correlations less significant,
similar to the
correlation between lensing potentials and the moving baryons responsible for the Ostriker-Vishniac/kinetic SZ effect \cite{OstVis86}.
Therefore, we ignore noise contributions from secondary effects associated with lensing extraction based on polarization estimators alone.

\begin{figure}[t]
\centerline{\psfig{file=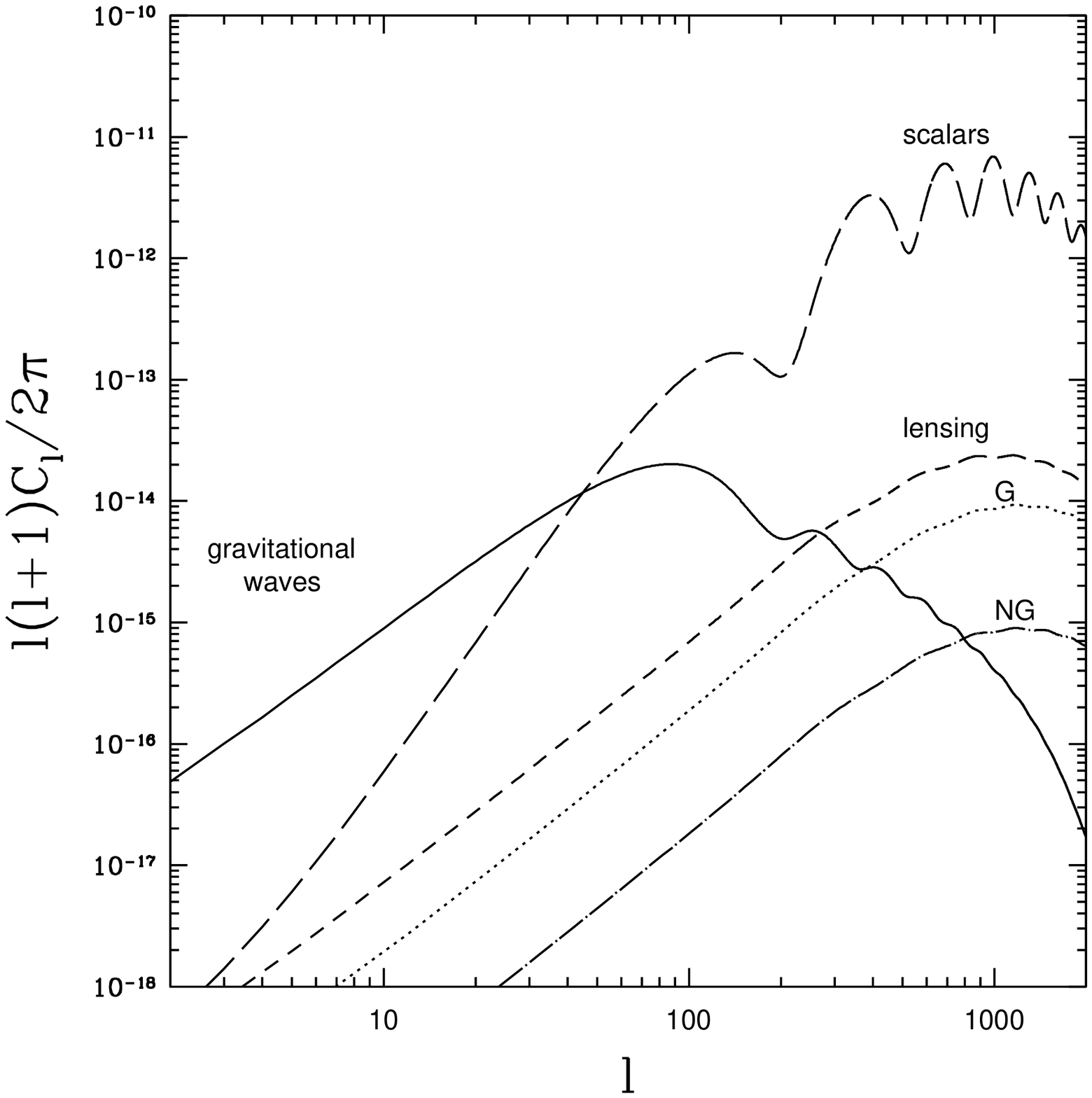,width=3.5in}
\psfig{file=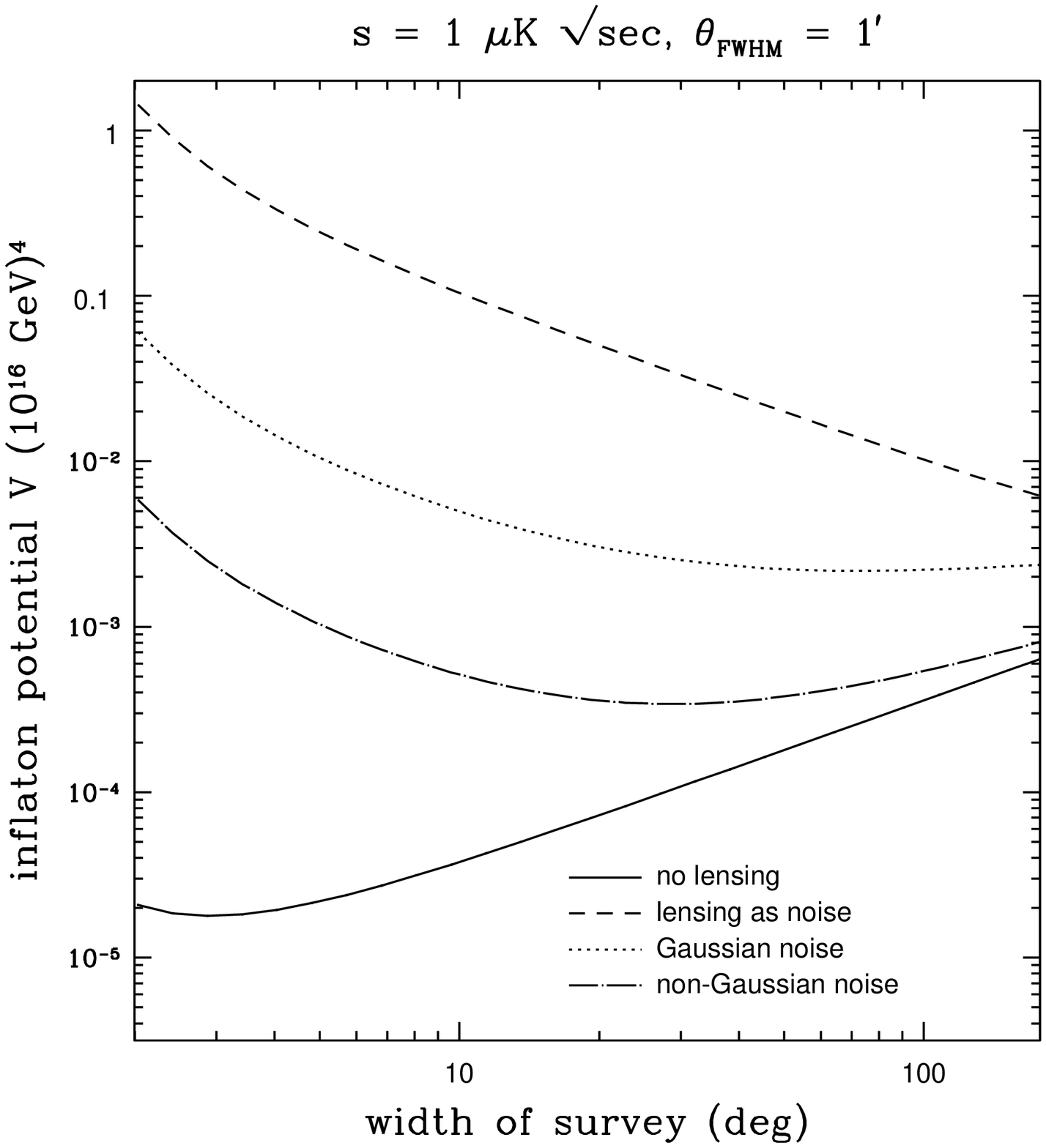,width=3.5in}}
\caption{{\it Left:} Various contributions to the CMB polarization. The
     long-dashed curve shows the dominant polarization signal in the
     gradient, or E-mode, component due to scalar perturbations while the solid
     line shows the curl, or B-mode, polarization signal from the
     gravitational-wave background assuming an
   inflationary energy scale of $2.3 \times 10^{16}$ GeV, consistent with the
limit based on COBE. The short-dashed curve shows the contribution to the
B-mode power spectrum due to gravitational lensing.  The dotted curve labeled
'G' is the residual on this contribution due to Gaussian noise
when lensing deflections are estimated from a temperature map with a
resolution of 1 arcmin, 
a sensitivity of 1 $\mu$K $\sqrt{\rm sec}$ and a sky coverage of 80\%.
The dot-dashed curves labeled
'NG' is the residual if the dominant Gaussian noise is removed as part of this
extraction leaving only the non-Gaussian noise of equation (\ref{E:FNG}).
{\it Right}: Minimum inflation potential observable at
     $1\sigma$ as a function of survey width for a one-year
     experiment. The solid curve shows results assuming no lensed E-mode
contribution to the B-mode map
while the short-dashed curve shows results including the effects of lensing as
a noise contribution.
	The dotted and dot-dashed curves correspond to the lensing
subtraction described in the left panel. The order of magnitude improvement
in the inflationary energy scales
that can be probed confusion-free with CMB polarization data suggests the
importance of investigating ways to
remove the noise-bias associated with the lensing extraction based on the
squared power spectrum.}
\label{fig:inflation}
\end{figure}

In general, the CMB data allow high signal-to-noise ratio reconstruction of the lensing effect. We illustrate the extent to which 
Planck's frequency cleaned temperature maps can be used as a lensing experiment in figure~(\ref{fig:convergence}). Here, we
show the angular power spectrum of convergence for sources at the last scattering, ie., CMB, and at a redshift of 1; the latter is
consistent with what is expected from weak lensing surveys via galaxy
ellipticity data. 
In the right panel, we show the cumulative signal-to-noise ratio for Planck based on the quadratic statistic discussed here and the
gradient-based statistic of \cite{SelZal99}. The reconstruction based on the
squared power spectrum has factor of $\sim$ 3 higher cumulative 
signal-to-noise than the one based on gradients in the temperature map. 
This can be understood by noting that the latter method
only extracts lensing from the first order expansion of the temperature in terms of the deflection angle, while significant contributions
also come from the second order term that is also probed by the proposed estimators. 

As shown in  figure~(\ref{fig:convergence}), due to Planck's beam of $\sim$ 5 to
10 arcmin the reconstruction of convergence becomes noise dominated at multipoles of
$\sim$ 1000, while one can probe lensing to much smaller
angular scales with galaxy information. The difference between the two curves
illustrate the importance of lensing reconstruction for
the polarization studies related to gravitational waves. Since CMB photons are
affected by potentials out to the last scattering surface,
 the resulting lensing effect is at least an order of magnitude higher than
the lensing effect on sources that are visible from the
large scale structure out to a redshift of a few. Thus, it is unlikely that
observations of the large scale structure lensing
can be used to subtract the lensing contribution to the B-mode
polarization sufficiently to permit a confusion free detection of
the polarization signal due to gravitational waves. 
We refer the reader to \cite{Kesetal02} for a further discussion on this
application. 

We can extend the discussion in  \cite{Kesetal02} to consider one interesting possibility. As discussed earlier,
one can make a significant improvement in the lensing reconstruction when the
Gaussian noise bias associated with the temperature squared power spectrum is removed.
In such a scenario, the noise contribution associated with
the lensing extraction is determined by the non-Gaussianities which are usually
an order of magnitude below the dominant Gaussian
noise contribution.  Such a Gaussian noise bias removal allows an extremely
high signal-to-noise extraction of the lensing signal
even for experiments with poor sensitivities that are not optimized for the
lensing reconstruction. 
As shown in figure~(\ref{fig:convergence}), for
Planck, the removal of the dominant Gaussian noise leads to detection of the
lensing angular power spectrum with a cumulative
signal-to-noise ratio of $\sim$ 700, which is roughly a factor of 20 above the
level expected when the dominant Gaussian noise contribution
is not removed. Such an extraction not only improves the effectiveness of
cosmological studies involving lensing of the CMB \cite{Hu02}, 
but also allows a high signal-to-noise separation of the lensed E-mode
contribution to the B-mode polarization map of the
CMB. This improves the confusion-free limit on the inflationary energy scale
attainable with CMB polarization
experiments by another order of magnitude \cite{Kesetal02}. 

Following \cite{Kesetal02}, we illustrate the importance of noise-bias removal
in figure~(\ref{fig:inflation}). 
Here, we compare various contributions to the power spectra of CMB
polarization, including the
lensed E-mode contribution to the B-mode polarization map. We also show the
residual lensing contribution
to the B-mode map after the lensing effect has been largely removed using an
estimator for lensing deflections.  The estimator employs
a CMB temperature map with a resolution of 1 arcmin and a sensitivity of 1
$\mu$K $\sqrt{\rm sec}$.
The curve labeled $NG$ is the residual contribution when the lensing
extraction is carried out to the noise level 
determined by the non-Gaussian noise contribution, assuming that the dominant
Gaussian noise contribution can be removed during the lensing
extraction. As shown, such a removal provides an order of magnitude
improvement over the residual from the dominant
Gaussian contribution.

In the right panel of figure~(\ref{fig:inflation}), 
we show limits on the inflationary energy scale based on lensing
extraction with the remaining residuals depicted in the left panel of the same
figure. We refer the reader to \cite{Kesetal02} for details related to this
calculation. For comparison, we also show the
inflationary energy scale limit 
when the lensing contribution is ignored (only instrumental noise)
and the limit when the lensing
contribution is treated as an additional
contribution to the noise associated with the B-mode map. As illustrated in
these figures, 
the potential removal of the Gaussian noise associated with the lensing
reconstruction reduces by an order of magnitude the confusion-free upper
bound on the gravitational wave signal in the B-mode polarization.
Such an improvement motivates the development of practical schemes to remove
the noise-bias associated with the dominant Gaussian
part of the temperature squared power spectrum. In future work, we will return
to this issue and will investigate ways to achieve the required level of noise
removal.

\section{Summary}
\label{sec:summary}

We have discussed the four-point correlation function, or the trispectrum in
Fourier space,
of CMB temperature and polarization anisotropies due to the weak 
gravitational lensing effect by intervening large scale structure.
The trispectrum contains information related to the lensing deflection angle,
which for a given configuration of the trispectrum is captured primarily by the
diagonal of the quadrilateral that defines the trispectrum in Fourier space.
We have devised a statistic
involving the power spectrum of squared temperatures that captures this
information from the diagonal.
This provides an observational approach to extracting the power spectrum of
deflection angles induced by the 
weak gravitational lensing effect on the CMB. 
The method extends previous suggestions on the use of the squared
temperature power spectrum
to extract the lensing-secondary correlation power spectrum by probing the
bispectrum formed by the lensing
effect. The technique presented here is not new and has been discussed
previously in the literature by
\cite{Hu01b}. We extend this work by providing a full derivation of the
trispectrum terms both
due to lensing and lensing-secondary correlations, and by providing a detailed
discussion of the filtering required to minimize the dominant noise component
due to Gaussian
temperature fluctuations that impairs the extraction of lensing information.

We also extend previous discussions on the trispectrum 
and associated weak lensing reconstruction from CMB data by calculating
in detail contributions to the noise in the reconstruction.  Previous work
discussed the dominant Gaussian noise, but we include in addition
terms associated with non-Gaussian contributions due to lensing alone
and the correlation of the
lensing effect with other foreground secondary anisotropies in the CMB such
as the ISW effect and the SZ effect.
When the SZ effect is removed from the temperature maps using its
spectral dependence, we find these additional non-Gaussian noise contributions
to be an order of magnitude lower than the dominant Gaussian noise. When SZ is
not removed, the extraction of lensing information is
significantly suppressed; we strongly advise the use of multiple frequencies
in upcoming small angular
scale CMB experiments, if there is an interest in using data from such
experiments to
conduct a weak lensing study. The lensing extraction provides a high
signal-to-noise ratio 
detection of the lensing deflection power spectrum. For the Planck surveyor,
with frequency-cleaned temperature
maps, detection will have a cumulative signal-to-noise of $\sim$ 35, while
certain techniques
in the literature \cite{SelZal99}, involving the gradients of the temperature
map, only give a cumulative signal-to-noise
ratio of $\sim$ 10.  Since the extraction with Planck is severely limited by
its beam of 5 to 10 arcmin,
we strongly recommend higher resolution observations at the arcminute scale.
If the noise-bias associated with the
dominant Gaussian part of the temperature squared power spectrum is removed,
Planck can extract the lensing power spectrum with temperature data alone
with a cumulative signal-to-noise ratio around $\sim$ 700.
With information from polarization, the signal-to-noise ratio associated with
lensing can be further increased by another factor of a few.
The order of magnitude improvement in the lensing separation 
motivates the development of practical methods that can remove the dominant
Gaussian noise contribution to the squared temperature power spectrum.
In general, our study suggests that future CMB observations can be used as a
weak lensing
experiment akin to current attempts involving lensing reconstruction from
galaxy ellipticity data. This opportunity will help to achieve an ultimate
challenge for CMB experiments, a confusion-free direct
detection of the gravitational wave signature from inflation.

\acknowledgments
We thank Wayne Hu and Marc Kamionkowski for useful discussions. This work was
supported in part by
NSF AST-0096023, NASA NAG5-8506, and DoE DE-FG03-92-ER40701.  Kesden
acknowledges the support of an NSF Graduate Fellowship and AC
acknowledges support from the Sherman Fairchild Foundation.

\section{Appendix I. EE and $\Theta$E estimators}

Here we provide the appropriate formulae for deriving optimal filters for the
EE and $\Theta$E estimators of $C_l^{\phi\phi}$.

\subsection{EE estimator}

The Fourier transform of the squared E-mode map can be expressed as a
convolution of Fourier transforms of the E-mode map itself, analogous to
equation (\ref{eqn:cmbsqr}) of \S~\ref{sec:covariance}.
\begin{equation} \label{E:EEdef}
(EE)(\vecl) = \int \frac{d^2\vecl_1}{(2\pi)^2} E(\vecl_1) E(\vecl-\vecl_1)
\end{equation}
Defining the power spectrum for the squared E-mode map as $\langle EE(\vecl) EE(\vecl')\rangle=(2\pi)^2 \delta(\vecl+\vecl') C_l^{E^2E^2}$,
we write
\begin{eqnarray}
\langle EE(\vecl) EE(\vecl')\rangle = 
 \int \frac{d^2\vecl_1}{(2\pi)^2} \int \frac{d^2\vecl_2}{(2\pi)^2} 
\langle E(\vecl_1) E(\vecl-\vecl_1) 
E(\vecl_2) E(\vecl'-\vecl_2) \rangle
\end{eqnarray}
As before, this power spectrum will consist of a noise component due to
Gaussian fluctuations in the E-mode map and a non-Gaussian contribution due
to lensing given respectively by, 
\begin{equation}
C_l^{E^2E^2 \rm G}
 = 2 \int\frac{d^2\vecl_1}{(2\pi)^2} C_{l_1}^{EE \tot}C_{|\vecl-\vecl_1|}^{EE \tot} \, ,
\end{equation}
\begin{eqnarray}
&& C_l^{E^2E^2 \rm NG} = C_l^{\phi\phi} \left[\intl{1} \Big\{
C_{l_1}^{EE} \vecl \cdot \vecl_1 
+C_{|\vecl-\vecl_1|}^{EE} \vecl \cdot (\vecl-\vecl_1) \Big\} \cos 2 (\varphi_{\vecl_1}+ \varphi_{\vecl-\vecl_1}) \right]^2
\nonumber \\
&+&\intl{1} \intl{2} \Big\{ C_{|\vecl_1+\vecl_2|}^{\phi\phi} \left[C_{l_1}^{EE} \vecl_1 + C_{l_2}^{EE} \vecl_2 \right] \cdot (\vecl_1 + \vecl_2)
\left[C_{|\vecl+\vecl_2|}^{EE} (\vecl+\vecl_2) - C_{|\vecl-\vecl_1|}^{EE} (\vecl-\vecl_1) \right] \cdot (\vecl_1 + \vecl_2) \nonumber \\
&& \quad \quad \times \cos 2(\varphi_{\vecl-\vecl_1} -\varphi_{\vecl+\vecl_2})  \cos 2(\varphi_{\vecl_1} +\varphi_{\vecl_2}) \nonumber \\
&+&C_{|\vecl-\vecl_1+\vecl_2|}^{\phi\phi} \left[C_{|\vecl-\vecl_1|}^{EE}
(\vecl-\vecl_1) + C_{l_2}^{EE} \vecl_2 \right] \cdot (\vecl-\vecl_1+\vecl_2)
\left[C_{|\vecl+\vecl_2|}^{EE}
(\vecl+\vecl_2 ) - C_{l_1}^{EE} \vecl_1 \right] \cdot (\vecl-\vecl_1+\vecl_2)
\nonumber \\
&& \quad \quad \times \cos 2(\varphi_{\vecl_1} -\varphi_{\vecl+\vecl_2})  \cos 2(\varphi_{\vecl-\vecl_1} +\varphi_{\vecl_2}) \Big\} \, .
\end{eqnarray}
In an analysis similar to that of \S~\ref{sec:covariance} and
\S~\ref{sec:EBtri}, we find that the optimal filter in the presence of the
Gaussian noise term above is
\begin{equation}
W(\vecl, \vecl_1) = \frac{\left[C_{l_1}^{EE} \vecl \cdot \vecl_1 
+C_{|\vecl-\vecl_1|}^{EE} \vecl \cdot (\vecl-\vecl_1)\right]\cos 2 (\varphi_{\vecl_1} +\varphi_{\vecl-\vecl_1})}{
2C_{l_1}^{EE \tot} C_{|\vecl-\vecl_1|}^{EE \tot}} \, .
\end{equation}
Using this filter, the dominant Gaussian noise $N_l$ is given by 
\begin{eqnarray}
N_l^{-1} = \intl{1} \frac{\left[C_{l_1}^{EE} \vecl \cdot \vecl_1 
+C_{|\vecl-\vecl_1|}^{EE} \vecl \cdot (\vecl-\vecl_1)\right]^2 \cos^2 2 (\varphi_{\vecl_1} + \varphi_{\vecl-\vecl_1})}{
2C_{l_1}^{EE \tot} C_{|\vecl-\vecl_1|}^{EE \tot} } \, ,
\end{eqnarray}
while the excess noise due to the additional terms in the E-mode trispectrum is
\begin{eqnarray} \label{E:NlEENG}
N_l^{\rm NG} &=& N_l^2 
 	\intl{1} \intl{2} \Big\{ W(\vecl, \vecl_1) W(-\vecl, \vecl_2)
\nonumber \\
&\times&C_{|\vecl_1+\vecl_2|}^{\phi\phi} \left[C_{l_1}^{EE} \vecl_1 + C_{l_2}^{EE} \vecl_2 \right] \cdot (\vecl_1 + \vecl_2)
\left[C_{|\vecl+\vecl_2|}^{EE} (\vecl+\vecl_2) - C_{|\vecl-\vecl_1|}^{EE} (\vecl-\vecl_1) \right] \cdot (\vecl_1 + \vecl_2) \nonumber \\
&& \quad \quad \times \cos 2(\varphi_{\vecl-\vecl_1} -\varphi_{\vecl+\vecl_2})  \cos 2(\varphi_{\vecl_1} +\varphi_{\vecl_2}) \nonumber \\
&+&C_{|\vecl-\vecl_1+\vecl_2|}^{\phi\phi} \left[C_{|\vecl-\vecl_1|}^{EE}
(\vecl-\vecl_1) + C_{l_2}^{EE} \vecl_2 \right] \cdot (\vecl-\vecl_1+\vecl_2)
\left[C_{|\vecl+\vecl_2|}^{EE}
(\vecl+\vecl_2 ) - C_{l_1}^{EE} \vecl_1 \right] \cdot (\vecl-\vecl_1+\vecl_2)
\nonumber \\
&& \quad \quad \times \cos 2(\varphi_{\vecl_1} -\varphi_{\vecl+\vecl_2})  \cos 2(\varphi_{\vecl-\vecl_1} +\varphi_{\vecl_2}) \Big\} \, .
\end{eqnarray}

\subsection{$\Theta$E estimator}
Here we derive the optimal filter and accompanying noise for the $\Theta$E
estimator of $C_l^{\phi\phi}$.  The following equations (\ref{E:cmbEdef})
through (\ref{E:NlcmbENG}) for this estimator are
entirely analogous to equations (\ref{E:EEdef}) through (\ref{E:NlEENG}) of
the preceding subsection of this Appendix. An additional complication here is
that, unlike in the $EB$ estimator
where $E$ and $B$ are uncorrelated due to parity,
$\theta$ and $E$ are correlated with a power
spectrum defined by $C_l^{\theta E}$ below. This produces an additional term
through the lensing effect in addition to that involving
correlations between $\theta$ and $E$ separately.
First, we expand in Fourier space the product of $\theta$ and $E$ maps,
\begin{equation} \label{E:cmbEdef}
\cmb E(\vecl) = \frac{1}{2}\int \frac{d^2\vecl_1}{(2\pi)^2} \left[\cmb(\vecl_1) E(\vecl-\vecl_1) +
E(\vecl_1) \cmb(\vecl-\vecl_1) \right] \, ,
\end{equation}
to obtain
\begin{eqnarray}
\langle \cmb E(\vecl) \cmb E(\vecl')\rangle = \frac{1}{4} 
 \int \frac{d^2\vecl_1}{(2\pi)^2} \int \frac{d^2\vecl_2}{(2\pi)^2} 
\langle \left[\cmb(\vecl_1) E(\vecl-\vecl_1) +E(\vecl_1) \cmb(\vecl-\vecl_1) \right] 
\left[\cmb(\vecl_2) E(\vecl'-\vecl_2) +E(\vecl_2) \cmb(\vecl'-\vecl_2) \right] \rangle \, ,
\end{eqnarray}
and 
\begin{equation}
C_l^{(\cmb E)^2 \rm G}
 = \frac{1}{4}\int\frac{d^2\vecl_1}{(2\pi)^2} \left[2C_{l_1}^{\cmb \tot}C_{|\vecl-\vecl_1|}^{EE \tot}+2C_{l_1}^{EE \tot}C_{|\vecl-\vecl_1|}^{\cmb \tot}
+4C_{l_1}^{\cmb E \tot}C_{|\vecl-\vecl_1|}^{\cmb E \tot} \right] \, .
\end{equation}
The non-Gaussian part is
\begin{eqnarray}
&& C_l^{(\cmb E)^2 \rm NG} = \frac{1}{4}C_l^{\phi\phi} \Big\{ \intl{1} \left[
C_{l_1}^{\cmb E} \vecl \cdot \vecl_1 
+C_{|\vecl-\vecl_1|}^{\cmb E} \vecl \cdot (\vecl-\vecl_1) \right]
\left[1+ \cos 2 (\varphi_{\vecl_1}+ \varphi_{\vecl-\vecl_1}) \right]
\Big\}^2 \nonumber \\
&+&\frac{1}{4} \intl{1} \intl{2} \Big\{ C_{|\vecl_1+\vecl_2|}^{\phi\phi} \Big\{ \left[C_{l_1}^{\cmb} \vecl_1 + C_{l_2}^{\cmb} \vecl_2 \right] \cdot (\vecl_1 + \vecl_2)
\left[C_{|\vecl+\vecl_2|}^{EE} (\vecl+\vecl_2) - C_{|\vecl-\vecl_1|}^{EE} (\vecl-\vecl_1) \right] \cdot (\vecl_1 + \vecl_2)
\cos 2(\varphi_{\vecl-\vecl_1} -\varphi_{\vecl+\vecl_2})
\nonumber \\
&+&\left[C_{l_1}^{EE} \vecl_1 + C_{l_2}^{EE} \vecl_2 \right] \cdot
(\vecl_1 + \vecl_2)
\left[C_{|\vecl+\vecl_2|}^{\cmb} (\vecl+\vecl_2) - C_{|\vecl-\vecl_1|}^{\cmb} (\vecl-\vecl_1) \right] \cdot (\vecl_1 + \vecl_2)
\cos 2(\varphi_{\vecl_1} +\varphi_{\vecl_2}) \nonumber \\
&+&\left[C_{l_1}^{\cmb E} \cos 2(\varphi_{\vecl_1} +\varphi_{\vecl_2}) \vecl_1
+ C_{l_2}^{\cmb E} \vecl_2 \right] \cdot
(\vecl_1 + \vecl_2)
\left[C_{|\vecl+\vecl_2|}^{\cmb E} \cos 2(\varphi_{\vecl-\vecl_1} -\varphi_{\vecl+\vecl_2})
(\vecl+\vecl_2) - C_{|\vecl-\vecl_1|}^{\cmb E} (\vecl-\vecl_1) \right] \cdot (\vecl_1 + \vecl_2) \nonumber \\
&+&\left[C_{l_1}^{\cmb E} \vecl_1
+ C_{l_2}^{\cmb E} \cos 2(\varphi_{\vecl_1} +\varphi_{\vecl_2}) \vecl_2 \right]
\cdot (\vecl_1 + \vecl_2)
\left[C_{|\vecl+\vecl_2|}^{\cmb E} 
(\vecl+\vecl_2) - C_{|\vecl-\vecl_1|}^{\cmb E} \cos 2(\varphi_{\vecl-\vecl_1} -\varphi_{\vecl+\vecl_2}) (\vecl-\vecl_1) \right] \cdot (\vecl_1 + \vecl_2)
\Big\} \nonumber \\
&+&C_{|\vecl-\vecl_1+\vecl_2|}^{\phi\phi} \Big\{
\left[C_{|\vecl-\vecl_1|}^{\cmb E}
(\vecl-\vecl_1) + C_{l_2}^{\cmb E} \cos 2(\varphi_{\vecl-\vecl_1} +\varphi_{\vecl_2}) \vecl_2 \right] \cdot (\vecl-\vecl_1+\vecl_2)
\left[C_{|\vecl+\vecl_2|}^{\cmb E}
(\vecl+\vecl_2 ) - C_{l_1}^{\cmb E} \cos 2(\varphi_{\vecl_1} -\varphi_{\vecl+\vecl_2}) \vecl_1 \right] \cdot (\vecl-\vecl_1+\vecl_2)
\nonumber \\
&+&\left[C_{|\vecl-\vecl_1|}^{\cmb E} \cos 2(\varphi_{\vecl-\vecl_1} +\varphi_{\vecl_2})
(\vecl-\vecl_1) + C_{l_2}^{\cmb E} \vecl_2 \right] \cdot (\vecl-\vecl_1+\vecl_2)
\left[C_{|\vecl+\vecl_2|}^{\cmb E} \cos 2(\varphi_{\vecl_1} -\varphi_{\vecl+\vecl_2})
(\vecl+\vecl_2 ) - C_{l_1}^{\cmb E} \vecl_1 \right] \cdot
(\vecl-\vecl_1+\vecl_2) \nonumber \\
&+&\left[C_{|\vecl-\vecl_1|}^{EE} (\vecl-\vecl_1) + C_{l_2}^{EE} \vecl_2
\right] \cdot
(\vecl -\vecl_1 + \vecl_2)
\left[C_{|\vecl+\vecl_2|}^{\cmb} (\vecl+\vecl_2) - C_{l_1}^{\cmb} \vecl_1
\right] \cdot (\vecl -\vecl_1 + \vecl_2)
\cos 2(\varphi_{\vecl -\vecl_1} +\varphi_{\vecl_2})
\nonumber \\
&+&\left[C_{|\vecl-\vecl_1|}^{\cmb} (\vecl-\vecl_1) + C_{l_2}^{\cmb} \vecl_2
\right] \cdot
(\vecl -\vecl_1 + \vecl_2)
\left[C_{|\vecl+\vecl_2|}^{EE} (\vecl+\vecl_2) - C_{l_1}^{EE} \vecl_1
\right] \cdot (\vecl -\vecl_1 + \vecl_2)
\cos 2(\varphi_{\vecl +\vecl_2} -\varphi_{\vecl_1})
\Big\} \Big\} \, .\nonumber \\
\end{eqnarray}
The filter required to suppress the dominant Gaussian noise and extract
lensing information is
\begin{equation}
W(\vecl, \vecl_1) = \frac{\left[C_{l_1}^{\cmb E} \vecl \cdot \vecl_1 
+C_{|\vecl-\vecl_1|}^{\cmb E} \vecl \cdot (\vecl-\vecl_1)\right] \left[1 +
\cos 2 (\varphi_{\vecl_1} +\varphi_{\vecl-\vecl_1}) \right]}{C_{l_1}^{\cmb \tot} C_{|\vecl-\vecl_1|}^{EE
\tot} + C_{l_1}^{EE \tot} C_{|\vecl-\vecl_1|}^{\cmb \tot}
 + 2C_{l_1}^{\cmb E \tot} C_{|\vecl-\vecl_1|}^{\cmb E \tot}} \, ,
\end{equation}
and the associated Gaussian noise is
\begin{equation}
N_l^{-1} = \intl{1} \frac{\left[C_{l_1}^{\cmb E} \vecl \cdot \vecl_1 
+C_{|\vecl-\vecl_1|}^{\cmb E} \vecl \cdot (\vecl-\vecl_1)\right]^2 \left[1 +
\cos 2 (\varphi_{\vecl_1} +\varphi_{\vecl-\vecl_1}) \right]^2}{2C_{l_1}^{\cmb \tot} C_{|\vecl-\vecl_1|}^{EE
\tot} + 2C_{l_1}^{EE \tot} C_{|\vecl-\vecl_1|}^{\cmb \tot}
 + 4C_{l_1}^{\cmb E \tot} C_{|\vecl-\vecl_1|}^{\cmb E \tot}} \, ,
\end{equation}
while the non-Gaussian noise is
\begin{eqnarray} \label{E:NlcmbENG}
C_l^{(\cmb E)^2 \rm NG} &=& \frac{N_l^2}{4} 
 	\intl{1} \intl{2} W(\vecl, \vecl_1) W(-\vecl, \vecl_2)
\nonumber \\
&\times& \Big\{ C_{|\vecl_1+\vecl_2|}^{\phi\phi} \Big\{ \left[C_{l_1}^{\cmb} \vecl_1 + C_{l_2}^{\cmb} \vecl_2 \right] \cdot (\vecl_1 + \vecl_2)
\left[C_{|\vecl+\vecl_2|}^{EE} (\vecl+\vecl_2) - C_{|\vecl-\vecl_1|}^{EE} (\vecl-\vecl_1) \right] \cdot (\vecl_1 + \vecl_2)
\cos 2(\varphi_{\vecl-\vecl_1} -\varphi_{\vecl+\vecl_2})
\nonumber \\
&+&\left[C_{l_1}^{EE} \vecl_1 + C_{l_2}^{EE} \vecl_2 \right] \cdot
(\vecl_1 + \vecl_2)
\left[C_{|\vecl+\vecl_2|}^{\cmb} (\vecl+\vecl_2) - C_{|\vecl-\vecl_1|}^{\cmb} (\vecl-\vecl_1) \right] \cdot (\vecl_1 + \vecl_2)
\cos 2(\varphi_{\vecl_1} +\varphi_{\vecl_2}) \nonumber \\
&+&\left[C_{l_1}^{\cmb E} \cos 2(\varphi_{\vecl_1} +\varphi_{\vecl_2}) \vecl_1
+ C_{l_2}^{\cmb E} \vecl_2 \right] \cdot
(\vecl_1 + \vecl_2)
\left[C_{|\vecl+\vecl_2|}^{\cmb E} \cos 2(\varphi_{\vecl-\vecl_1} -\varphi_{\vecl+\vecl_2})
(\vecl+\vecl_2) - C_{|\vecl-\vecl_1|}^{\cmb E} (\vecl-\vecl_1) \right] \cdot (\vecl_1 + \vecl_2) \nonumber \\
&+&\left[C_{l_1}^{\cmb E} \vecl_1
+ C_{l_2}^{\cmb E} \cos 2(\varphi_{\vecl_1} +\varphi_{\vecl_2}) \vecl_2 \right]
\cdot (\vecl_1 + \vecl_2)
\left[C_{|\vecl+\vecl_2|}^{\cmb E} 
(\vecl+\vecl_2) - C_{|\vecl-\vecl_1|}^{\cmb E} \cos 2(\varphi_{\vecl-\vecl_1} -\varphi_{\vecl+\vecl_2}) (\vecl-\vecl_1) \right] \cdot (\vecl_1 + \vecl_2)
\Big\} \nonumber \\
&+&C_{|\vecl-\vecl_1+\vecl_2|}^{\phi\phi} \Big\{
\left[C_{|\vecl-\vecl_1|}^{\cmb E}
(\vecl-\vecl_1) + C_{l_2}^{\cmb E} \cos 2(\varphi_{\vecl-\vecl_1} +\varphi_{\vecl_2}) \vecl_2 \right] \cdot (\vecl-\vecl_1+\vecl_2) \nonumber \\
&& \quad \quad \times \left[C_{|\vecl+\vecl_2|}^{\cmb E}
(\vecl+\vecl_2 ) - C_{l_1}^{\cmb E} \cos 2(\varphi_{\vecl_1} -\varphi_{\vecl+\vecl_2}) \vecl_1 \right] \cdot (\vecl-\vecl_1+\vecl_2)
\nonumber \\
&+&\left[C_{|\vecl-\vecl_1|}^{\cmb E} \cos 2(\varphi_{\vecl-\vecl_1} +\varphi_{\vecl_2})
(\vecl-\vecl_1) + C_{l_2}^{\cmb E} \vecl_2 \right] \cdot (\vecl-\vecl_1+\vecl_2)
\left[C_{|\vecl+\vecl_2|}^{\cmb E} \cos 2(\varphi_{\vecl_1} -\varphi_{\vecl+\vecl_2})
(\vecl+\vecl_2 ) - C_{l_1}^{\cmb E} \vecl_1 \right] \cdot
(\vecl-\vecl_1+\vecl_2) \nonumber \\
&+&\left[C_{|\vecl-\vecl_1|}^{EE} (\vecl-\vecl_1) + C_{l_2}^{EE} \vecl_2
\right] \cdot
(\vecl -\vecl_1 + \vecl_2)
\left[C_{|\vecl+\vecl_2|}^{\cmb} (\vecl+\vecl_2) - C_{l_1}^{\cmb} \vecl_1
\right] \cdot (\vecl -\vecl_1 + \vecl_2)
\cos 2(\varphi_{\vecl -\vecl_1} +\varphi_{\vecl_2})
\nonumber \\
&+&\left[C_{|\vecl-\vecl_1|}^{\cmb} (\vecl-\vecl_1) + C_{l_2}^{\cmb} \vecl_2
\right] \cdot
(\vecl -\vecl_1 + \vecl_2)
\left[C_{|\vecl+\vecl_2|}^{EE} (\vecl+\vecl_2) - C_{l_1}^{EE} \vecl_1
\right] \cdot (\vecl -\vecl_1 + \vecl_2)
\cos 2(\varphi_{\vecl +\vecl_2} -\varphi_{\vecl_1})
\Big\} \Big\} \, . \nonumber \\
\end{eqnarray}

The trispectrum formed from two temperature and two E-mode terms will contain
an additional non-Gaussian contribution through the
coupling of the lensing deflection angle to secondary effects.  A similar
contribution occurs in the temperature trispectrum and was discussed at the end
of \S~\ref{sec:lensing}.  As in equation (\ref{eqn:trisec}), 
cumulants such as $\langle \cmb(\vecl) \cmb^{\rm s}(\vecl^{\prime})\rangle$
will vanish
as the secondary effects are decoupled from the primary fluctuations generated
at the surface of last scatter. However,
contributions come from the correlation between $\cmb^{\rm s}$ and the lensing
deflection $\phi$.
As before, contributions of equal importance come from both the first and
second order terms in $L$ written in
equation~(\ref{eqn:lfactor}).  We consider this additional non-Gaussian
contribution for the connected portion of a four-point function in Fourier
space
\begin{eqnarray}
\left< \tilde E(\bfl_1) \tilde E(\bfl_2) \cmb^\tot(\bfl_3)
       \cmb^\tot(\bfl_4)\right>_c &=& \Big< \left( E(\vecla) - \intl{1'} E(\vecla') \cos 2(\varphi_{\vecl_{1}^{\prime}} - \varphi_{\vecl_1})
L(\vecla,\vecla')\right) \nonumber \\
&& \quad \left(E(\veclb) - \intl{2'} E(\veclb')
\cos 2(\varphi_{\vecl_{2}^{\prime}} - \varphi_{\vecl_2})
L(\veclb,\veclb')\right) \cmb^\s(\veclc)  \cmb^\s(\vecld) \Big> \nonumber \\
&=& - C_{l_1}^{EE} \cos 2(\varphi_{\vecl_1} + \varphi_{\vecl_2})
\left< L(\veclb,-\vecla)\cmb^s(\veclc) \cmb^\s(\vecld) \right> 
- C_{l_2}^{EE} \cos 2(\varphi_{\vecl_1} + \varphi_{\vecl_2})
\left< L(\vecla,-\veclb) \cmb^s(\veclc)  \cmb^\s(\vecld) \right>  \nonumber \\
&& \quad + \intl{1'} C_{l_1'}^{EE}
\cos 2(\varphi_{\vecl_{1}^{\prime}} - \varphi_{\vecl_1})
\cos 2(\varphi_{\vecl_{1}^{\prime}} + \varphi_{\vecl_2})
\left< L(\veclb,-\vecla') 
L(\vecla,\vecla') \cmb^s(\veclc)  \cmb^\s(\vecld) \right>
\nonumber \\
\label{eqn:TEtrisec}
\end{eqnarray}
Contributions to the trispectrum from the first two terms come
through the second order term in $L$, with the two $\phi$ terms
coupling to $\cmb^{\rm s}$. In the last term, contributions come
from the first order term of $L$ similar to the pure
 lensing contribution to the trispectrum.

After some straightforward simplifications, we can write the
connected part of the trispectrum involving lensing-secondary coupling.  We
symmetrize with respect to the four indices by adding five terms corresponding
to the additional permutations of $(l_1,l_2,l_3,l_4)$ and multiplying by a
factor of 1/6.
\begin{eqnarray}    
&& T^{\cmb E\tot}(\bfl_1,\bfl_2,\bfl_3,\bfl_4) = \nonumber \\
&& \frac{1}{6}
C_{l_3}^{\len\s} C_{l_4}^{\len\s} \Big\{  C^\cmb_{l_1}
\cos 2(\varphi_{\vecl_1} + \varphi_{\vecl_2})
(\veclc \cdot \vecla) (\vecld \cdot \vecla) + 
 C^\cmb_{l_2} \cos 2(\varphi_{\vecl_1} + \varphi_{\vecl_2})
(\veclc \cdot \veclb) (\vecld \cdot \veclb) \nonumber \\
&& \quad \quad + \left[ \veclc \cdot (\vecla + \veclc) \right]
\left[ \vecld \cdot (\veclb +\vecld) \right] C^\cmb_{|\vecl_1 + \vecl_3|} 
\cos 2(\varphi_{\vecl_1 + \vecl_3} - \varphi_{\vecl_1})
\cos 2(\varphi_{\vecl_2 + \vecl_4} - \varphi_{\vecl_2}) \nonumber \\
&& \quad \quad +\left[ \vecld \cdot (\vecla + \vecld) \right]
\left[ \veclc \cdot (\veclb +\veclc) \right] C^\cmb_{|\vecl_1 + \vecl_4|}
\cos 2(\varphi_{\vecl_1 + \vecl_4} - \varphi_{\vecl_1})
\cos 2(\varphi_{\vecl_2 + \vecl_3} - \varphi_{\vecl_2}) + \, {\rm Perm.}
\Big\} \nonumber \\
\end{eqnarray}
Note that the first two terms come from the first and 
second term in equation~(\ref{eqn:TEtrisec}), 
while the last two terms are from the third term.


\begin{thebibliography}{99}
\frenchspacing
\bibitem{PeeYu70}
P. J. E. Peebles and J. T. Yu, \ApJ, {\bf 162}, 815 (1970);
R. A. Sunyaev and Ya.B. Zel'dovich, Astrophys. Space Sci. {\bf 7}, 3 (1970);
J. Silk, \ApJ, {\bf 151}, 459 (1968); see recent review by
W. Hu \& S. Dodelson, Ann. Rev. Astro. Astrop., in press, 
astro-ph/0110414 (2001).


\bibitem{Jugetal95}
\aut{Jungman}{G}, \aut{Kamionkowski}{M}, \aut{Kosowsky}{A} \amp
\aut{Spergel}{D.N},\refs{Cosmological-Parameter Determination with Microwave
Background Maps}{\PRD}{54}{1332}{1995}{astro-ph/9512139};
\aut{Bond}{J.R}, \aut{Efstathiou}{G} \amp \aut{Tegmark}{M},
       \refs{Forecasting Cosmic Parameter Errors from Microwave
        Background Anisotropy Experiments}
        {\MNRAS}{291}{L33}{1997}{astro-ph/9702100};
\aut{Zaldarriaga}{M}, \aut{Spergel}{D.N} \amp \aut{Seljak}{U},
       \refs{Microwave Background Constraints on Cosmological Parameters}
        {\ApJ}{488}{1}{1997}{astro-ph/9702157};
\aut{Eisenstein}{D.J}, \aut{Hu}{W} \amp \aut{Tegmark}{M},
       \refs{Cosmic Complementarity: Joint Parameter Estimation from
        CMB Experiments and Redshift Surveys}
       {\ApJ}{518}{2}{1999}{astro-ph/9807130}


\bibitem{Miletal99} A. D. Miller et al., \ApJL\ {\bf 524}, L1 (1999);
        P. de Bernardis et al.,
     Nature {\bf 404}, 955
     (2000); S. Hanany et al., \ApJL\ {\bf 545}, L5 (2000);
        N. W. Halverson et al., astro-ph/0104489.

\bibitem{Coo02} see, for example, review by A. Cooray, 2002, in ``2001 Coral Gables
Conference: Cosmology and Particle Physics'', eds. B.
       Kursunoglu \& A. Perlmutter; American Institute of Physics Conference Proceedings, astro-ph/0203048

\bibitem{Blaetal87}
A. Blanchard and J. Schneider, \AsAs, 184, 1 (1987);
A. Kashlinsky, \ApJ, 331, L1 (1988);
E. V. Linder, \AsAs, 206, 1999, (1988);
S. Cole and G. Efstathiou, \MNRAS, 239, 195 (1989);
M. Sasaki, \MNRAS, 240, 415 (1989);
K. Watanabe and K. Tomita, \ApJ, 370, 481 (1991);
M. Fukugita, T. Futumase, M. Kasai and E. L. Turner, \ApJ, 393, 3 (1992);
L. Cayon, E. Martinez-Gonzalez and J. Sanz, \ApJ, 413, 10 (1993);
U. Seljak, \ApJ, 463, 1 (1996)	

\bibitem{SpeGol99}
        D. N. Spergel and D. M. Goldberg, \PRD {\bf 59}, 103001 (1999);
        D. M. Goldberg and D. N. Spergel, \PRD, {\bf 59}, 103002 (1999);
        A. Cooray and W. Hu, \ApJ, {\bf 534}, 533 (2000).
        M. Zaldarriaga and U. Seljak, \PRD, {\bf 59}, 123507 (1999);
        H. V. Peiris and D. N. Spergel, \ApJ, {\bf 540}, 605 (2000).

\bibitem{Zal00}
	F. Bernardeau, A\&A, {\bf 324}, 15 (1997).
        M. Zaldarriaga, \PRD, {\bf 62}, 063510 (2000).

\bibitem{Hu01}
	W. Hu, \PRD, {\bf 64}, 083005 (2001)

\bibitem{Benetal00}
	K. Benabed, F. Bernardeau and L. van Waerbeke, \PRD, {\bf 63}, 043501 (2001); J. Guzik, U. Seljak and M. Zaldarriaga, \PRD, {\bf 62} 043517 (2000); 

\bibitem{SacWol67}
        R. K. Sachs and A. M. Wolfe, \ApJ, {\bf 147}, 73 (1967).

\bibitem{SunZel80}
        R. A. Sunyaev and Ya. B. Zel'dovich, \MNRAS, {\bf 190}, 413 (1980).

\bibitem{Hu00}
        W. Hu, \PRD, {\bf 62}, 043007  (2000).

\bibitem{Coo02b}
	A. Cooray, \PRD, {\bf 65}, 063512 (2002).

\bibitem{SelZal99}
        U. Seljak \& M. Zaldarriaga, \PRL, {\bf 82}, 2636 (1999); 
	M. Zaldarriaga and U. Seljak, \PRD, {\bf 58} 023003 (1998).	

\bibitem{Cooetal00}
        A. Cooray, W. Hu and M. Tegmark, \ApJ, {\bf 540}, 1 (2000).

\bibitem{Kesetal02}
        Kesden, M., Cooray, A., Kamionkowski, M. 2002, PRL submitted, astro-ph/0202434;
	Knox, L. \& Song, Y.-S. 2002, PRL submitted, astro-ph/0202286.          

\bibitem{KamKosSte97} M. Kamionkowski, A. Kosowsky, and
     A. Stebbins, \PRL\ {\bf 78}, 2058 (1997);
        U. Seljak and M. Zaldarriaga, \PRL\ {\bf 78}, 2054 (1997).


\bibitem{ZalSel98} M. Zaldarriaga and U. Seljak, \PRD\ {\bf
     58}, 023003 (1998).

\bibitem{KamKos99}  See, for example, M. Kamionkowski and A. Kosowsky,
     Ann. Rev. Nucl. Part. Sci., {\bf 49}, 77 (1999).

\bibitem{Hu01b}
	W. Hu, \ApJ, {\bf 557}, 79 (2001).

\bibitem{HuOka01}
	W. Hu and T. Okamoto, ApJ submitted, astro-ph/0111606 (2001).





\bibitem{Kai92}
      N. Kaiser, \ApJ, {\bf 388}, 286 (1992);
      N. Kaiser, \ApJ, {\bf 498}, 26 (1998);
	M. Bartelmann \& P. Schneider, Physics Reports in press, astro-ph/9912508 (2000).


\bibitem{Lim54}
        D. Limber, \ApJ, {\bf 119}, 655 (1954).


\bibitem{Bar80}
	J. M. Bardeen \PRD {\bf 22} 1882 (1980).


\bibitem{EisHu99}
        D. J. Eisenstein \& W. Hu, \ApJ, {\bf 511}, 5 (1999).

\bibitem{BunWhi97}
       E. F. Bunn \& M. White, \ApJ, {\bf 480}, 6 (1997).

\bibitem{ViaLid99}
        P. T. P. Viana \& A. R. Liddle, MNRAS, {\bf 303}, 535 (1999).

\bibitem{Pee80}
        P. J. E. Peebles,  The Large-Scale Structure of the
        Universe, Princeton: Princeton Univ. Press (1980).

\bibitem{PeaDod96}
        J. A. Peacock and S. J. Dodds MNRAS, {\bf 280} L19 (1996).

\bibitem{KamSpe94}
	Marc Kamionkowski and David N. Spergel, \ApJ, {\bf 432}, 7 (1994).

\bibitem{Val00}
	P. Valageas, A\&A, {\bf 354}, 767 (2000).


\bibitem{Kno95} L. Knox, \PRD\ {\bf 52}, 4307 (1995).       

\bibitem{Coo01}
	A. Cooray, \PRD, {\bf 64},  043516 (2001).

\bibitem{SelZal96}
        U. Seljak \& M. Zaldarriaga
        \ApJ, {\bf 469}, 437 (1996).



\bibitem{Hu00b}
	 W. Hu, \ApJ, {\bf 529}, 12 (2000).  	




\bibitem{OstVis86}
        J. P. Ostriker and E. T. Vishniac, \ApJ, {\bf 306}, L51 (1986);
	E. T. Vishniac, \ApJ, {\bf 322}, 597 (1987);
	S. Dodelson and J. M. Jubas, \ApJ, {\bf 439}, 503 (1995);
        G. Efstathiou, in Large Scale Motions in the Universe. A
Vatican Study Week, ed. V. C. Rubin \& G. V. Coyne (Princeton: Princeton
University Press), 299 (1988);
        A. H. Jaffe and M. Kamionkowski, \PRD, {\bf 58}, 043001 (1998);

\bibitem{Hu02}
	Hu, W. \PRD, {\bf 65}, 023003 (2002).



\end{thebibliography}
\end{document}